# Depletant-DNA Induces Chirality in the Condensed Domains of Lyotropic Liquid Crystals

**Zeba Afia Hasan[a+], Elizabeth Adeogun[a+], Harold Hatch[b], Jacob Monroe[a], Karthik Nayani[a*]**

[a] Department of Chemical Engineering, University of Arkansas, Fayetteville, AR, USA, [b] National Institute of Standards and Technology, Gaithersburg, MD, USA.
+ co-first authorship

**We report on the phase behavior of solutions of a chromonic liquid crystal, disodium cromoglycate (DSCG) resulting from the presence of DNA that acts as a depletant. We show that DNA (both single and double stranded) can induce phase condensation of DSCG at volume fractions a thousand-fold lower than a commonly studied depletant of similar size, namely polyethylene glycol. Twist angle characterization via polarized optical microscopy reveals DNA introduces macroscopic chirality within the condensed phase, despite being present only in the continuous phase- as confirmed by imaging of fluorescently tagged DNA. UV-vis absorption spectroscopy reveals that the introduction of chirality reduces the absorption coefficient strongly, indicating that induced chirality leads to considerable increase in the average lengths of DSCG aggregates. We also show that DNA at concentrations (~nmol/L) typical for a post amplification event (such as PCR) can induce an isotropic-biphasic phase transition of DSCG enabling a fast optical method to report the presence of DNA. Overall, these findings establish DNA as exceptionally efficient depletants that regulate both phase condensation and chirality transfer in lyotropic liquid crystals and thereby providing a simple optical platform for detection of amplified DNA.**

## Introduction

Nature utilizes chirality as a tool for self-assembly and as a design principle in many clever ways. For instance, the selective recognition of L-amino acids by cellular enzymes arises from the chiral specificity of biomolecular interactions (*1*). Despite the prevalence of scientific inquiry on chirality dating back to the 1700s, there remain several unanswered questions regarding the organization and self-assembly introduced due to chirality within the body (*2*). Additionally, in crowded biological environments of cells, depletion forces play a central role in driving organization and phase behavior by promoting the condensation and ordering of larger structures and increasing the configurational entropy of the system (*3, 4*).

DNA is an interesting molecule in the context of both processes described above as it possesses backbone chiral centers, helical architecture, anisotropic shape, and tunable length scales, all of which can strongly influence collective assembly processes (*5*). Consequently, DNA represents a particularly intriguing depletant, not only because of its geometry, but also because its chirality may direct the organization of surrounding mesoscopic structures. Understanding how chiral depletants mediate biomolecular organization, liquid-liquid phase separation, and self-assembly in crowded biological systems remains an important and yet unexplored central biological question (*6-18*).

A facile platform to systematically explore the role of a chiral depletant (such as DNA) is provided by lyotropic chromonic liquid crystals (LCLCs). They are composed of planar aromatic cores that lack stereogenic centers making the molecules mirror symmetric (*19, 20*). In aqueous solutions, these molecules undergo reversible self-assembly processes driven by π-π stacking forming elongated rod-like aggregates (*19-24*). At sufficiently high concentrations, these rod-like aggregates self-align to form liquid crystalline phases (*19, 25*). Prior work has revealed that non-adsorbing crowding agents, such as polyethylene glycol (PEG), can induce phase separation in a LCLC, namely, disodium cromoglycate (DSCG) into coexisting isotropic and nematic phases at a volume fraction around ~ Ø ~$5 \times 10^{-3}$ (*26*). Previous studies have also demonstrated chiral induction in DSCG solutions via a) the inclusion of small molecule chiral dopants in the condensed phase (*27*) and b) emergence of twist (equal proportions of right and left handedness) from the interplay between elastic distortions and boundary conditions in spatially confined droplets (*28-31*) and tactoids (*26-28, 32*). Our initial motivation of using DNA was to explore the role of the shape of depletant in the phase behavior of DSCG. Despite decades of research on depletion (*6-8, 10, 11, 16, 33-36*), systems wherein both the depletant (DNA) and the particles associating via depletion (DSCG) are both rod-like are scarce (*16*). However, we find here that the chirality of the depletant has a much more prominent role in the phase behavior of the condensed phase as compared to the shape of the depletant. Via this work, we address two fundamental questions: i) Can a depletant impart chirality to the condensed phase while remaining in the continuous phase? ii) How does the induced chirality influence the self-assembly of the condensed phase? More broadly, we note that DNA-induced chirality within a condensed phase has important biological consequences for cellular organization and processes (*37, 38*).

In addition to providing fundamental insight into how chirality of a depletant can influence phenomenology pertaining to liquid-liquid phase separation, our study also hints at a basis of an approach for rapid identification of low concentrations of DNA. Typical concentrations of DNA following a polymerase chain reaction (PCR) are in the nanomolar range (*39*). LCs have been shown to be excellent optical transducers of presence of small molecules in a range of contexts. We show that nanomolar DNA can induce phase condensation of DSCG and that this phase change is also sensitive to number of base pairs of DNA-thereby permitting the development of a facile optical method to detect sequenced DNA.

## RESULTS AND DISCUSSION

**Phase Behavior of DSCG in presence of PEG and DNA:** A DSCG molecule (shown in Figure 1A) has a disc-like shape with an aromatic core and ionic peripheral groups (*40, 41*). These disc-like molecules form rod-like aggregates via $\pi$-$\pi$ stacking interactions as depicted in Figure 1A depending on temperature, concentration, presence of depletants etc. (*41, 42*). The phase behavior of DSCG in the absence of any depletants as a function of temperature and concentration is presented in Figure 1B. For all experiments reported in this work, the starting concentration of DSCG is 9.5 wt% (0.095 mass fraction) and we infer from Figure 1B that at this concentration, we expect to find DSCG in its isotropic phase at 22°C. We observe in Figure 1C, that in the presence of PEG, 9.5 wt% DSCG has undergone a phase change from isotropic (dark under crossed-polarizers) to the biphasic regime (birefringent nuclei in a dark background under crossed-polarizers). Figure 1C reports the volume fractions of PEG (of varying Mw) needed to nucleate the first nematic domains of DSCG (9.5 wt%) Consistent with prior reports, addition of 2 kDa PEG at a volume fraction of $5.26\times10^{-2}$ to 9.5 wt% DSCG results in the emergence of the first birefringent domains, as shown in Figure 1C (*40*). Systematic study of PEGs of varying sizes, namely, 2 kDa, 4 kDa, 10 kDa and 35 kDa (having radius of gyration of 2.45 nm, 4.29 nm, and 8.5 nm respectively) reveal the emergence of birefringent domains (and hence the biphasic region) at volume fractions of $5.26\times10^{-2}$, $1.82\times10^{-2}$, $1.63\times10^{-2}$, and $1.54\times10^{-3}$, respectively as shown in Figure 1C. Crucially, we note that the volume fraction of PEG needed to induce the phase condensation is of the order ~ $10^{-2}$.

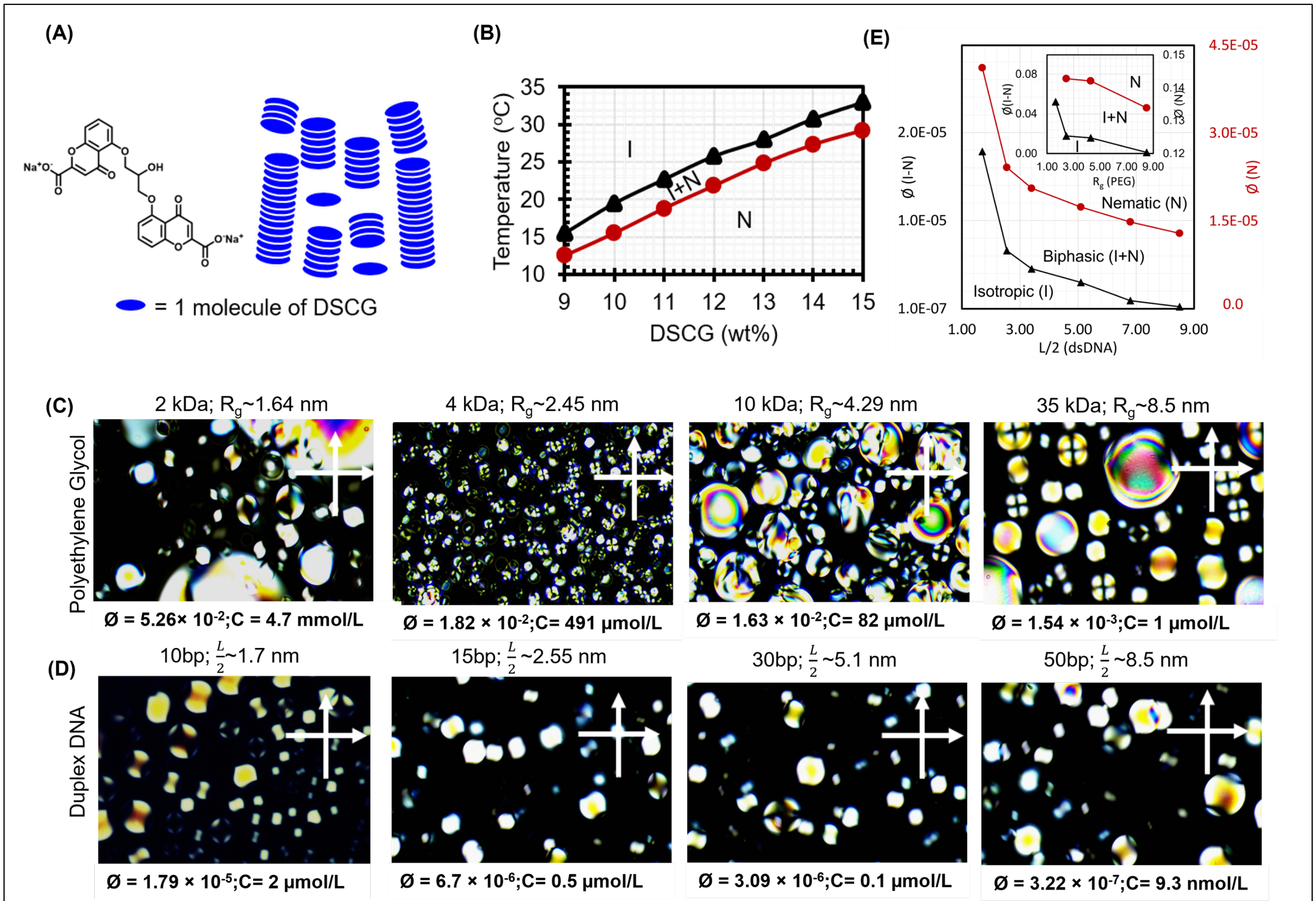

**Figure 1: Characterization and phase behavior of liquid crystalline phases of DSCG in the presence of varying sizes of dsDNA and PEG.** (A) Molecular structure and schematic of self-assembly of DSCG in its nematic phase. (B) Phase diagram of DSCG as a function of concentration and temperature. Black curve denotes the measured isotropic (I)-to-biphasic (I–N) phase boundary, while red curve denotes the biphasic (I-N)-to-nematic (N) phase boundary. (C) Cross-polarized optical microscopy images of 9.5 wt% DSCG in presence of threshold PEG volume fractions required to induce phase condensation with varying radii of gyration ($R_g$) of PEG; isotropic regions appear dark, whereas nematic regions appear bright due to birefringence. (D) Cross-polarized optical microscopy images of 9.5 wt% DSCG in presence of the threshold dsDNA volume fractions required to induce phase condensation. dsDNA lengths (L/2) are similar to $R_g$ of PEG in the panel above. (E) Phase behavior of 9.5 wt% DSCG as a function of the dsDNA length (L/2). Black curve and red curve denote the I to I-N and I–N to N phase boundaries respectively, at 22°C. Inset: Corresponding phase behavior of 9.5 wt% DSCG as a function of the PEG's $R_g$ at 22°C.

We performed experiments to characterize the phase behaviour of DSCG with dsDNA as depletant and these results are presented in Figure 1D. Surprisingly, the volume fraction of DNA needed to induce phase condensation of DSCG was three orders of magnitude lower than that of PEG of similar size ($\frac{L}{2}$ ~ radii of gyration, $R_g$). From Figure 1D we note that varying base pairs (10, 15, 30, and 50 bp) of dsDNA oligomers induce phase condensation of 9.5 wt% DSCG at volume fractions of $1.79\times10^{-5}$, $6.7\times10^{-6}$, $3.09\times10^{-6}$, and $3.22\times10^{-7}$ respectively. The sizes of DNA used correspond closely to PEG used in Figure 1C ($\frac{L}{2}$ ~ $R_g$). Remarkably, the longest dsDNA (50 bp) triggers the isotropic-to-biphasic transition at an extremely low volume fraction of Ø = $3.22\times10^{-7}$- approximately four orders of magnitude below the PEG volume fraction (Ø = $1.54\times10^{-3}$) required to induce the same transition.

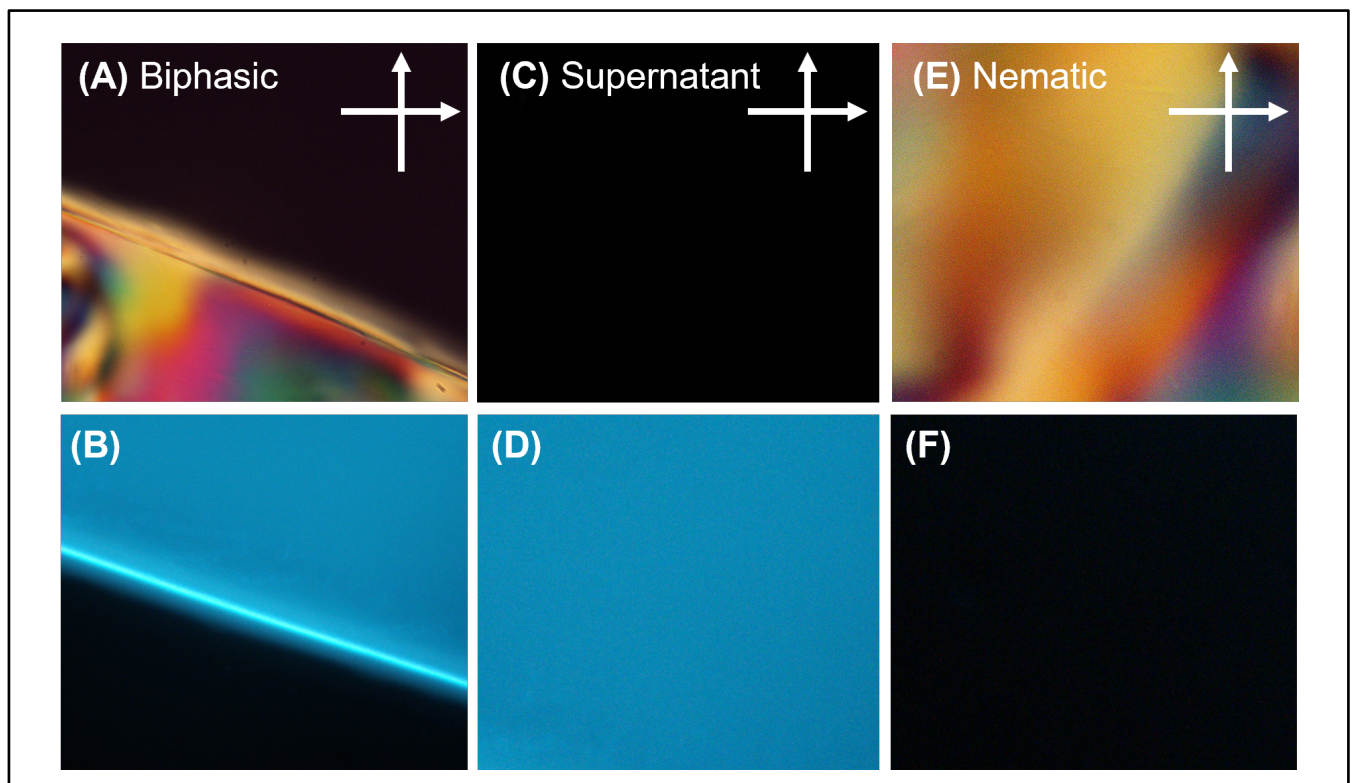

**Figure 2: Evaluating dsDNA as a non-adsorbing depletant.** (A, C, E) Cross-polarized optical microscopy images of the biphasic, isotropic supernatant, and nematic phases, respectively. (B, D, F) Corresponding fluorescence microscopy images of the same field of view shown in (A), (C), and (E).

We also studied the phase behavior of DSCG using fluorescently labeled dsDNA (5′ modification with fluorescein that has emission wavelength of 517 nm) to verify that dsDNA is acting as a non-adsorbing depletant. Mixtures of 10 bp dsDNA at Ø = $2\times10^{-4}$ and 9.5 wt% DSCG (isotropic, as shown in Figure S1) were placed in a vial and left to phase separate. After a few minutes of equilibration, the mixture undergoes well-defined macroscopic phase separation into a dense condensed phase and a dilute supernatant phase (as shown in Figure S2). Figure 2 presents imaging of a given sample in: cross-polarized optical microscopy (top panel) and fluorescence microscopy (bottom panel) of the biphasic (Figure 2A and 2B), supernatant (Figure 2C and 2D), and nematic regimes (Figure 2E and 2F). In the biphasic image shown in Figure 2A we can see that the isotropic part is on the top portion of the image while the nematic part is in the bottom portion. Comparing with the corresponding fluorescence image of Figure 2B, we can clearly see that the signal is confined only to the top portion and therefore the isotropic phase. Line scan intensity along the diagonal of the image in Figure 2B is presented in Figure S3. From Figure S3 we note that the grayscale fluorescence intensity in the nematic part of the image is close to zero whereas the isotropic intensity averages around 110. Similar observations were made for the supernatant and nematic domains as well. The supernatant phase remains completely dark under crossed polarizers (Figure 2C), indicating the absence of liquid-crystalline order. The corresponding fluorescence image (Figure 2D) exhibits strong fluorescence, confirming that DNA preferentially resides in the isotropic phase. In contrast, the dense nematic phase is uniformly birefringent as shown in Figure 2E, whereas the corresponding fluorescence image in Figure 2F shows no fluorescence intensity. Grayscale intensity of the line scan of the supernatant in Figure 2D also averages around 110 whereas the grayscale intensity of the line scan of the nematic phase in Figure 2F is once again close to zero. These observations demonstrate, based on limit of detection of fluorescence microscopy, dsDNA is excluded from the condensed nematic DSCG domains and remains in the surrounding isotropic (continuous) region, consistent with the behavior of a non-adsorbing depletant.

**Role of Depletant Shape:** We sought to explore if the rod-like shape of DNA can explain its efficiency as depletant in comparison to its spherical counterpart, PEG. Characteristics of the depletant, such as its shape, can play an important role in the phase behavior of mixtures (*43*). For instance, plates are depleted more strongly by infinitely thin rods than spheres(*6, 7, 44, 45*). In case of binary mixtures of large and small colloidal hard spheres, the depletion potential in the lowest order of density is given by: $W(h) = -3k_BTØ_s\frac{R}{\sigma}(1-\frac{h}{\sigma})^2$, while the depletion potential of big spheres depleted by small rods is $W(h) = -\frac{2}{3}k_BTØ_r\frac{L}{D}\frac{R}{D}(1-\frac{h}{L})^3$ (where $R$ is the radius of big spheres, $\sigma$ is the diameter of the small sphere depletants, $L$ is the length of rod-shaped depletants, $D$ is the diameter of rod shaped depletants, $Ø_s$ *and* $Ø_r$ are the volume fraction of small spheres and rod-shaped depletants respectively) (*6, 16, 46*). When the rod diameter equals the sphere diameter ($D = \sigma$), rods produce a stronger interaction at equal volume fraction when $L/D > 4.5$. If the rod length equals the sphere diameter ($L = \sigma > D$), rods produce a stronger interaction when $L/D > 2.12$. However, these previous theories cannot account for approximately three orders-of-magnitude lower concentration required for dsDNA than PEG to induce phase condensation in our system. Additionally, theoretical descriptions of systems wherein both the depletant and the particles associating via depletion being rod-shaped are scarce (*13*).

To examine the theoretical predictions in this regime and in the size-ranges relevant for our system, we consider i) Analytical calculations of rods and spheres depleting plates and ii) Monte Carlo simulations of rods and spheres depleting other rods. Full details are provided in Materials and Methods section. From Figure 3A, we note that rods of the same volumes as spheres, but with a higher aspect ratio, lead to stronger association. For instance, for an aspect ratio of ~5 our calculations reveal that rods should induce a depletion force about ~1.5 times stronger than spheres of the same volume. This situation maps to experimentally comparing 2kDa PEG with 30 bp DNA in Figure 1. From Figure 1 we note that 30 bp DNA induces phase condensation of DSCG at volume fractions 4 orders of magnitude lower than 2kDa PEG. It is therefore likely that a far greater depletion force is induced by DNA (in comparison to PEG) as opposed to the factor of 1.5 predicted by the computations. Additionally, rods with total lengths matching the diameters of spheres (resulting in a lower rod volume) lead to reduced attractive interactions (Figure 3A). This is, however, largely due to larger volumes of the spheres in comparison with rods (explained in detail below). In Figure 3B, we consider potentials of mean force (PMFs) computed from Monte Carlo simulations of rods comparable to DSCG in size and aspect ratio being depleted by DNA or PEG (*D=L*) of different sizes. Our simulations reveal that PEG generates stronger contact free energy (difference in PMF value in contact versus infinitely separated) than DNA for a similar size (*D=l*), but this is largely due to the larger volume occupied by a sphere compared to a cylinder of the same size. When normalized for their respective volume fractions, we find that DNA induces stronger contact potentials compared to PEG. For instance, in Figure 3B, comparing PEG ($D$= 4.9) and DNA ($l$=5.1) at $h$=2.5, we note that the PMF generated by PEG is about twice as strong as that of DNA. However, after normalizing with their respective volumes, we find that DNA ($l$=5.1) generates 2 times larger PMF than PEG ($D$=4.9). While such PMFs only consider pairwise DSCG interactions, higher-order interactions are not expected to significantly alter trends when comparing DNA and PEG depletants. Our experimental observations in Figure 1B (second panel) however reveal that DNA (same size as the computation) induces the biphasic transition at a volume fraction three orders of magnitude lower than PEG of the same size, indicating the PMF induced by DNA is far greater than the factor of 2 obtained by the computations. Thus, our computations cannot explain the 1000-fold more efficient condensation induced by dsDNA by shape-factor arguments alone. Therefore, we consider other influences, such as chirality of the depletant, to explain the observed phenomenology.

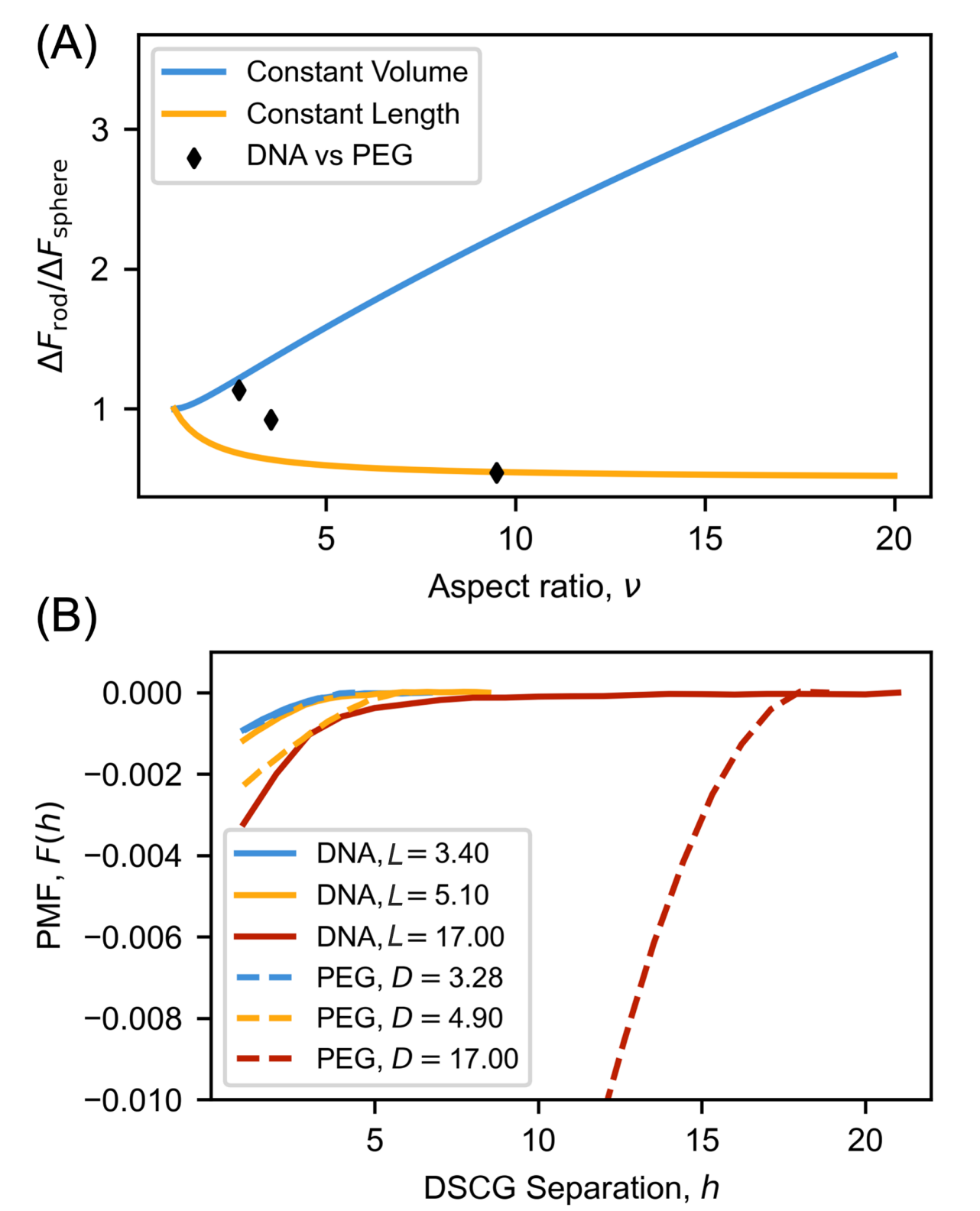

**Figure 3: Theoretical calculations of depletion interactions.** (A) Contact free energies between large plates are calculated for both rod and sphere depletants, with the ratio shown as a function of aspect ratio ($\nu = (L + D)/D$) of the rods. The blue curve considers the case of constant volume between rods and spheres, orange considers rods with constant total length matching the diameter of a sphere, and points being a comparison of PEG and DNA of similar sizes considered in experiments. (B) Potentials of mean force calculated from Monte Carlo simulations for two parallel rods (at the average stack size for DSCG columns) being depleted by a single DNA or PEG, each approximated by a hard rod or hard sphere, respectively.

**Induction of Chirality in the Condensed Phase via a Chiral Depletant:** To explore whether chirality of the depletant plays a role in the organization of the condensed phase of DSCG (i.e., the nematic domains of the biphasic region), we characterize the director configuration within tactoids nucleated by PEG (10 kDa, having radii of gyration of 4.29 nm) and dsDNA (30 bp, having L/2 ~ 5.1 nm). Tactoids are the first nematic domains that nucleate from the isotropic phase when LCLCs undergo an isotropic to biphasic transition. In a conventional untwisted tactoid, the director field remains mirror-symmetric about the tactoid axis as presented in the schematic of Figure S4A. An untwisted tactoid when imaged with cross-polarized optical microscopy has a dark central region. Upon slightly uncrossing the polarizers the transmitted intensity through the tactoid-center increases because the analyzer is no longer exactly orthogonal to the polarization state of the light exiting the tactoid. Consequently, the tactoid-center becomes brighter. We observe this exact phenomenology when imaging tactoids of DSCG nucleated by PEG (2 kDa). In Figure 4A we see a tactoid with its long axis parallel to a polarizer present with a dark central region. Upon slightly uncrossing the polarizers (P-A angle, $\gamma$ = 100°) the center of the tactoid becomes brighter consistent with the expectation of an untwisted tactoid. We plot the transmitted intensity as a function of the angle between the polarizer and analyzer ($\gamma$) in Figure 4E (blue data points). We note that the minimum intensity in Figure 4E is observed when $\gamma$ = 90° and the plot is symmetric with respect to minima. This plot is in quantitative agreement with the expected transmitted intensity profile of an untwisted tactoid ($\tau$ = 0). Similar results were obtained with 35kDa PEG as the depletant (Figure S5).

Strikingly, we observe qualitatively different phenomenology within a DSCG tactoid when dsDNA is the depletant. In Figure 4F we note that the center of the tactoid is not completely dark when $\gamma$ = 90° (dsDNA is the depletant). Upon uncrossing the polarizers ($\gamma$ = 100°), we can clearly see the center of the tactoid is darker. Under Mauguin conditions, where the optical phase retardation greatly exceeds the total director rotation ($\frac{\pi|\Delta n|d}{\lambda} \gg \tau$), the polarization of light adiabatically follows the rotating liquid-crystal director as it propagates through the sample thickness (*47-49*). Figure 4J presents the transmitted intensity (blue points and blue line) as a function of the angle between the polarizer and analyzer ($\gamma$) with dsDNA (30 bp) as the depletant. We note that the minima occur at $\gamma$ = 100° (experimental data was obtained with rotating the analyzer at 10° increment). The blue scattered datapoints represent experimental data whereas the blue trendline corresponds to the fit obtained using the transmittance equation for a twisted nematic liquid crystal cell-

$$T = cos^2\beta - \frac{\tau}{2\delta}\sin(2\delta)\cos(2\beta)\left[\frac{\tau}{\delta}\tan(\delta) + \tan(2\beta)\right] \ldots\ldots\ldots [1]\ (50)$$

where $\delta = \sqrt{\tau^2 + (\frac{\pi\Delta n d}{\lambda})^2}$, $\beta = \gamma - \tau$, and $\tau$ is the total twist angle. Birefringence, $\Delta$n = -0.02 (*51*) and wavelength, $\lambda$ = 550 nm. The corresponding angular dependence of transmitted intensity shows a shift of the extinction minimum away from 90°, confirming a 12.7° twist (*28, 52*). Figure S4B represents the director field inside a twisted tactoid rotated by 12.7°. Average values of twist obtained within DSCG tactoids in the presence of dsDNA as depletant was 12 ± 3°. For the sample thickness d = 20 μm, the Mauguin parameter is, $\frac{\pi|\Delta n|d}{\lambda\tau} = \frac{\pi\times|-0.02|\times 20\times 180}{550\times 12.72\pi}$ ≈10.3≫1, confirming that the Mauguin limit is satisfied and so the light polarization adiabatically follows the twisted director field.

Further, we also characterized the condensed nematic regions when confined within rectangular capillaries. Rectangular capillaries lacks the spatial confinement-related effects observed in tactoids (*26*) due to the relatively flat surfaces far from the edges. Figure 4C shows condensed DSCG phase in presence of PEG (2 kDa) depletant within rectangular capillaries under crossed polarizers. When we uncrossed the polarizers, as shown in Figure 4D, we see that the center of the capillary becomes brighter, confirming that PEG does not generate twist in the condensed LC phase. The red datapoints and trendline in Figure 4E correspond to the experimental and theoretical transmitted intensity respectively as a function of $\gamma$ for the DSCG + PEG confined to a rectangular capillary. Similar to the tactoids, we observe a minima at $\gamma = 90°$ confirming the lack of twist within the sample.

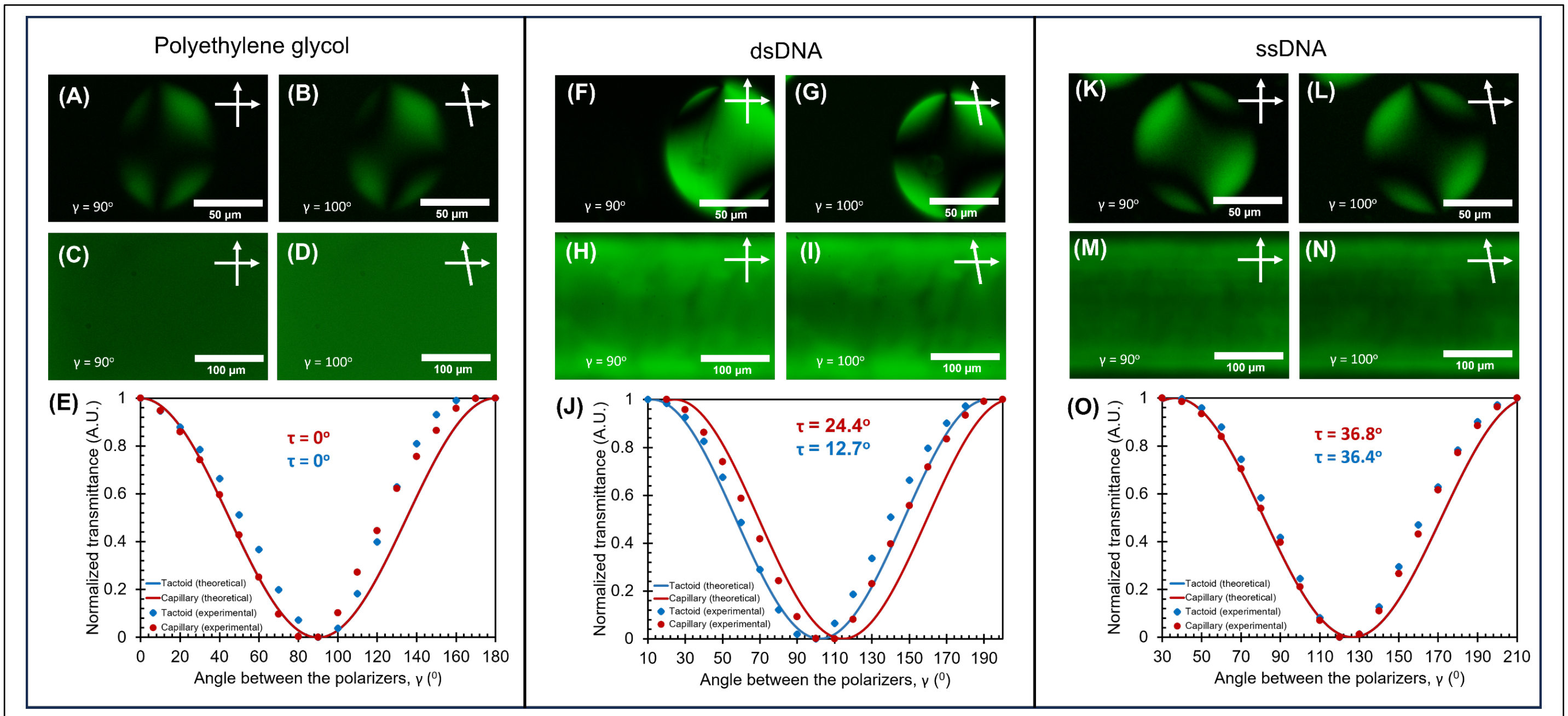

**Figure 4: Optical evidence of chirality transfer and enhanced phase condensation efficiency of chiral depletants.** (A-D, F-I, K-N) Cross-polarized ($\gamma$ =90°) and slightly uncrossed ($\gamma$ =100°) optical microscopy images of 9.5 wt% DSCG with (A-B) 1.95 µM dsDNA (rectangular capillary), (C-D) 1.95 µM dsDNA tactoids, (F-G) PEG (rectangular capillary), (H-I) PEG tactoids, (K-L) 1.95 µM ssDNA (rectangular capillary), and (M-N) 1.95 µM ssDNA tactoids. (E, J, O) Transmitted light intensity through the center of the rectangular capillary (red trendline) and tactoid (blue trendline) as a function of the angle difference ($\gamma$) between the polarizer and analyzer. Experimental data points are fits to equation 1. Throughout this manuscript, we refer to nmol/L as nM and µmol/L as µM .

On the contrary, Figure 4I ($\gamma$ = 100°) shows that in the presence of dsDNA, the center of the capillary containing DSCG exhibits a pronounced reduction in transmitted intensity when the polarizers are uncrossed from Figure 4H ($\gamma$ = 90°). The transmitted intensity data presented in Figure 4J shows that, unlike the PEG-DSCG system, the DSCG-dsDNA system in the rectangular capillary exhibits an extinction minimum at $\gamma \neq 90°$, indicating a rotation of the polarization state of the light by a twisted director field. Fitting the transmitted intensity in equation 1 yields a twist angle of 24.4°. Average values of twist obtained in the rectangular capillaries containing DSCG in the presence of dsDNA as depletant was 24 ± 4°. The persistence of twist within the rectangular capillaries demonstrates that twist is not restricted to confinement induced by the tactoids.

Further, to highlight the critical role played by chirality, we also characterized the behavior of DSCG when using single-stranded DNA (ssDNA) as a depletant. Similar to PEG, ssDNA is a flexible spherical coil while also possessing chirality akin to dsDNA. Figures 4 K through L show DSCG-tactoid nucleated in the presence of ssDNA (100 bases, having radii of gyration of 4.43 nm), while Figures 4M through N show a condensed DSCG phase in presence of ssDNA confined to a rectangular capillary. Notably, we observe the qualitatively similar phenomenology when ssDNA and dsDNA are depletants. In Figure 4L we note that, the center of the tactoid is not completely dark when $\gamma$ = 90° (ssDNA is the depletant). Upon uncrossing the polarizers ($\gamma$ = 100°), we can clearly see the center of the tactoid becomes noticeably darker. Figure 4O presents the experimental (blue datapoints) and theoretical (blue trendline) transmitted intensity as a function of the angle between the polarizer and analyzer ($\gamma$) with ssDNA (100 bases) as the depletant. We note that the intensity minima occur at $\gamma$ = 120°. The blue scattered datapoints once again represent experimental data whereas the blue trendline corresponds to the fit to equation 1, yielding a twist angle of 36.4°. Average values of twist obtained within DSCG tactoids in the presence of ssDNA as depletant was 36 ± 6°. Rectangular capillaries containing DSCG and ssDNA exhibited the same qualitative behavior as those containing dsDNA. Figure 4N ($\gamma$ = 100°) shows that upon uncrossing the polarizers from Figure 4M ($\gamma$ = 90°), the center of the capillary exhibits significant reduction in transmitted intensity. The transmitted intensity profile ( red datapoints correspond to experimental data and red trendline corresponds to theoretical fit to equation 1) in Figure 4O exhibits an extinction minimum shifted away from $\gamma$ = 90°, consistent with a twisted director field that rotates the polarization state of the transmitted light. Fitting the transmitted intensity in equation 1 yields a twist angle of 36.8°. Average values of twist obtained in the rectangular capillaries containing DSCG in the presence of dsDNA as depletant was 36 ± 8°. While the twist angles induced by ssDNA are larger in comparison to those induced by dsDNA, we note that at the concentration of 1.95 µM (as was maintained in Figure 4) the volume fraction of dsDNA is an order of magnitude lower than ssDNA. Even slight increases in the dsDNA concentrations (5 µM) led to a transition from biphasic to a complete nematic phase. Twist angles of DSCG in capillaries with 5 µM dsDNA led to an increase in the twist angle measured at around ~ 35 ± 6° as shown in Figure S6.

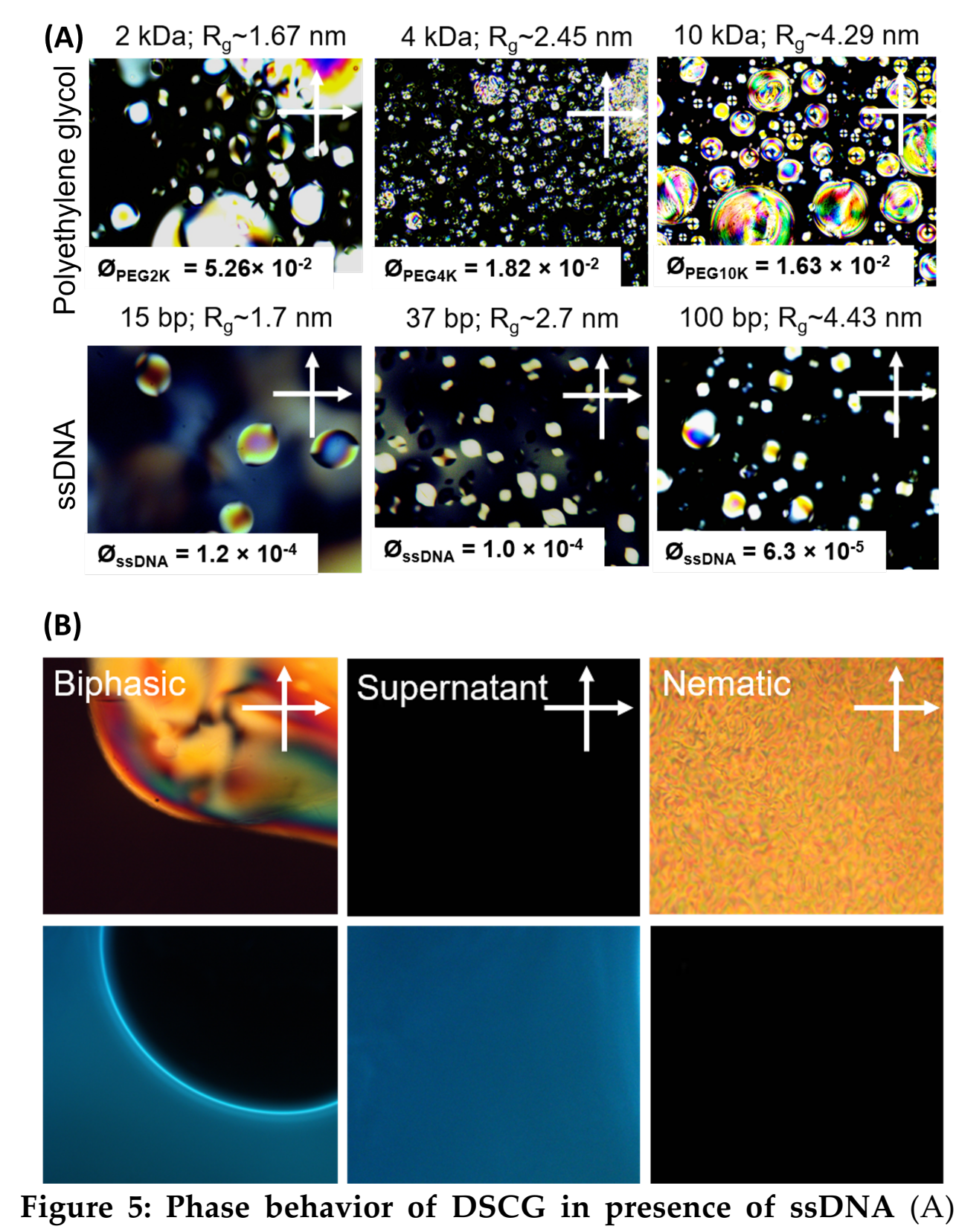

**Figure 5: Phase behavior of DSCG in presence of ssDNA** (A) Phase behavior of DSCG in presence of ssDNA and PEG showing ssDNA induce phase transition at much lower volume fraction than PEG despite having the same spherical shape. (B) Cross-polarized (top) and fluorescent (bottom) images of the supernatant, coexistence, and dense phases for DSCG in presence of ssDNA. The corresponding ssDNA concentrations for the 15-, 37-, and 100-base oligonucleotides are approximately 9.7, 2, and 0.3 µM, respectively.

Across all samples (n=52) the measured twist angles, for both dsDNA and ssDNA (of varying concentrations 5 nM- 5$\mu$M), are always positive, indicating a consistently right-handed twist throughout the condensed DSCG phase. This observation is consistent with the transfer of chirality from the right-handed DNA helices despite their exclusion from the condensed nematic domains. This behavior contrasts sharply with the confinement-induced twisted structures in LCLCs wherein both left- and right-handed tactoids form in equal proportions (*26*). Consequently, although individual tactoids are chiral, the ensemble remains racemic and exhibits no macroscopic homochirality (*26*). Likewise, in rectangular capillaries, twist nucleates from the curved boundaries through saddle-splay elasticity, generating director configurations of opposite handedness with equal probability in the absence of an intrinsic chiral field (*53*). In contrast, the exclusive observation of positive twist angles in the present study demonstrates that DNA is transferring its chirality to DSCG aggregates, imparting a unique handedness and establishing a globally homochiral condensed phase.

To also test the depleting ability of ssDNA, we compared the volume fractions of ssDNA needed to induce condensation of DSCG in comparison with PEG of similar size. Figure 5A presents a quantitative comparison between PEG and ssDNA of similar radius of gyration in term of volume fraction required to induce phase condensation in DSCG. In the top panel we find again that the volume fractions of PEG (2 kDa, 4 kDa, and 10 kDa) required to nucleate the first nematic domains of DSCG (9.5 wt%) whereas the bottom panel presents the phase behavior of DSCG with ssDNA as depletant. ssDNA base pairs are chosen to have radii of gyration comparable to those of the PEG depletants. We find that ssDNA induces phase condensation in DSCG at volume fractions at least two orders of magnitude lower than PEG, despite both depletants having a similar spherical conformation and comparable radius of gyration, differing primarily in the chiral backbone of ssDNA. We note that ssDNA oligomers with radius of gyration of 1.7 nm, 2.7 nm, and 5.5 nm can induce phase condensation of 9.5 wt% DSCG at volume fractions of $1.2\times10^{-4}$, $1.0\times10^{-4}$, and $6.3\times10^{-5}$ respectively. Whereas the top panel shows that PEG of similar radius of gyration require volume fraction of $5.26\times10^{-2}$, $1.82\times10^{-2}$, and $1.63\times10^{-2}$ respectively to induce the same phase condensation in DSCG.

To confirm that ssDNA also behaves as a non-adsorbing depletant, we examined its spatial distribution following phase separation. A mixture of 9.5 wt% DSCG and fluorescently labeled ssDNA of 100 bases (5′ fluorescein modification) at $Ø = 6.3 \times 10^{-5}$ was allowed to phase separate in a vial for approximately 10 min. Figure 5B presents cross-polarized optical microscopy (top panel) and fluorescence microscopy (bottom panel) images of the same field of view in the biphasic, supernatant, and nematic regimes. We report qualitatively similar phase behavior of DSCG as with dsDNA as the depletant. In the biphasic regime, the isotropic phase occupies the upper portion of the image, while the nematic phase is located in the lower portion. The corresponding fluorescence image reveals that the fluorescence signal is confined exclusively to the isotropic region. Similar behavior is observed for the supernatant and nematic phases. The supernatant remains completely dark under crossed polarizers, indicating the absence of liquid-crystalline order, whereas the corresponding fluorescence image exhibits strong fluorescence, confirming that ssDNA preferentially resides in the isotropic phase. In contrast, the condensed nematic phase is uniformly birefringent, while the corresponding fluorescence image shows essentially no fluorescence. These observations, together with the dsDNA results, demonstrate that both dsDNA and ssDNA behave as non-adsorbing depletants, partitioning exclusively into the isotropic (continuous) phase while transferring chirality to the condensed DSCG phase across the phase boundary.

**Influence of Chirality on Aggregation Behavior of DSCG**

To understand the enhanced phase condensation efficiency of chiral depletants, we studied the effect of chirality on the length of DSCG aggregates via UV-vis absorption measurements. Such experiments on dilute solutions of LCLCs have been employed previously to deduce the average aggregation number via fits to law of mass action models (*54*). As higher concentrations of DSCG lead to saturation of the signal, we performed UV-vis absorption spectroscopy on 1 mM DSCG in the presence of PEG, dsDNA, and ssDNA of similar size and volume fractions. Figure 6 shows the UV-vis spectra of DSCG and DSCG containing 5 μM PEG (10 kDa), 25 μM dsDNA (30 bp), 2.25 μM ssDNA (100 bases). The molar absorption coefficient is calculated from optical density (OD) using the formula $\varepsilon = A/Cd$, where $A$ is the absorbance equivalent to OD, $C$ is the molar concentration of DSCG, and $d$ is path length (= 0.322 cm). All spectra retain the same overall shape (peak at 325 nm), indicating that the presence of either achiral or chiral depletants does not alter the characteristic 0.34 nm π-π stacking distance in DSCG aggregates. However, the absorption coefficient at the characteristic DSCG wavelength of 325 nm decreases systematically upon the addition of depletants. For aggregating species such as DSCG, a reduction in the absorption coefficient corresponds to an increase in the average aggregate length (*51*). We note from Figure 6 that adding PEG (2.25 μM) to 1mM DSCG corresponds to a slight decrease in the molar absorption coefficient, but a significant reduction is observed when same volume fractions of ssDNA and dsDNA are the depletants. PEG volume fractions needed to be 1000-fold larger (~ 5mM) to observe a similar decrease of the molar absorption coefficient of DSCG as compared to the chiral depletants (Figure 6). These results clearly show the prominent effect of chirality of the depletant in influencing the aggregation lengths of DSCG.

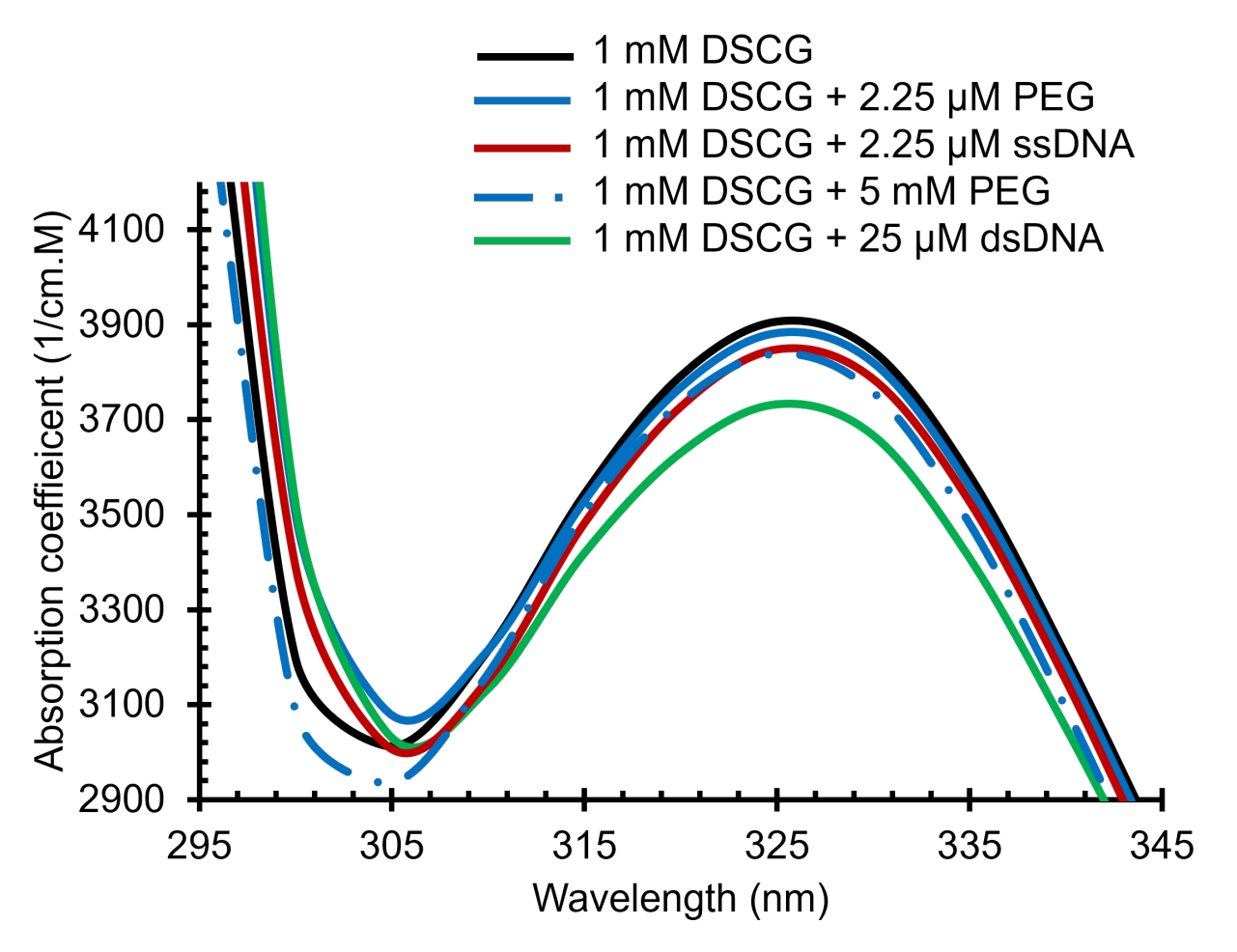

**Figure 6:** UV-vis spectra showing the molar absorption coefficients for 1 mM DSCG and 1mM DSCG with; PEG, dsDNA and ssDNA as depletants.

It has previously been computed that having a chiral dopant like l-alanine within the DSCG phase can lead to an increase in the aggregation energy and therefore larger aggregates (*55*). More recently, a theoretical description of chirality amplification in liquid crystals, wherein a weak chiral bias synchronizes fluctuating left- and right-handed conformations into a cooperative chiral state was shown (*56*). Consistent with this description, we hypothesize that the enhanced aggregation lengths we observe originates from chirality bias in the presence of DNA. In the absence of any chiral bias, DSCG aggregates are expected to populate weakly-twisted left- and right-handed aggregates with equal probability (*55*). Introducing right-handed DNA as a depletant appears to bias the dynamic equilibrium between left- and right-handed DSCG aggregates and effectively increasing the population of homochiral aggregates. We surmise that such cooperative chiral ordering increases the probability of homochiral interactions between neighboring aggregates (*56*), facilitating stacking and effectively, resulting in longer DSCG aggregates.

To quantify the increase in average DSCG aggregate length induced by the different depletants, we mapped the maximum absorption coefficient at 325 nm for PEG-, dsDNA-, and ssDNA-containing samples onto the calibration curve of pure DSCG with varying concentrations, as shown in Figure S7A. The resulting average aggregate lengths correspond to effective DSCG concentrations of 1.05, 1.12, and 1.30 mM for DSCG + PEG, DSCG + ssDNA, and DSCG + dsDNA systems respectively. To estimate the corresponding aggregate size distributions, we employed a law-of-mass-action model that assumes isodesmic aggregation (*54, 57, 58*). Using this model, Figure S7B shows the distribution of the relative volume fraction $X_N$ of aggregates containing $N$ molecules as a function of $N$ for the 4 experimental conditions described above, namely DSCG, DSCG + PEG, DSCG + ssDNA, and DSCG + dsDNA. From this $X_N(N)$ distribution, we compute an ensemble average value of $\langle X_N \frac{L}{D} \rangle$. In several theoretical treatments, including Onsager's description of the isotropic-nematic transition imply higher values of $\langle X_N \frac{L}{D} \rangle$ (with $X_N$ as an analogue for volume fraction) correspond to higher likelihood of the formation of a nematic phase (*24, 59*). We see from the plot in Figure S7B that the addition of dsDNA corresponds to a 68% increase in the value of Image (4.08 to 6.86) when compared to pure DSCG, whereas when compared to DSCG in the presence of PEG, the increment is 65%. These results demonstrate that chiral DNA depletants promote formation of longer DSCG aggregates when compared to PEG, providing a mechanistic basis for the exceptionally low DNA concentrations required to induce phase condensation.

## Rapid Detection of RPA-amplified DNA

This fortuitous discovery of LC phase transitions in the presence of unusually low volume fraction of DNA opens an avenue for optical reporting of DNA amplification. We explored the recombinase polymerase amplification (RPA) method to amplify targeted DNA fragments while using phase transitions in DSCG as an optical transducing mechanism. An RPA kit comprising of positive control primers, magnesium acetate ($Mg(OAc)_2$), and a target nucleic acid were mixed to form a 50 $\mu$L solution that was incubated at 37 °C for 20 minutes. Figure 7A shows a schematic of our setup; a 6 $\mu$L aliquot of the resulting product was subsequently added to 24 $\mu$L of 13 wt% DSCG to obtain a final concentration of 9.5 wt% DSCG. When imaged with cross-polarized optical microscopy, we found that the positive control sample is biphasic (Figure 7A). Interestingly, the mixture with the negative control sample (comprised of all the buffers and primers except for the target nucleic acid) resulted in the isotropic phase shown in Figure 7A. Figure 7B shows the gel electrophoresis results for positive and negative controls confirming the presence of DNA in the positive control.

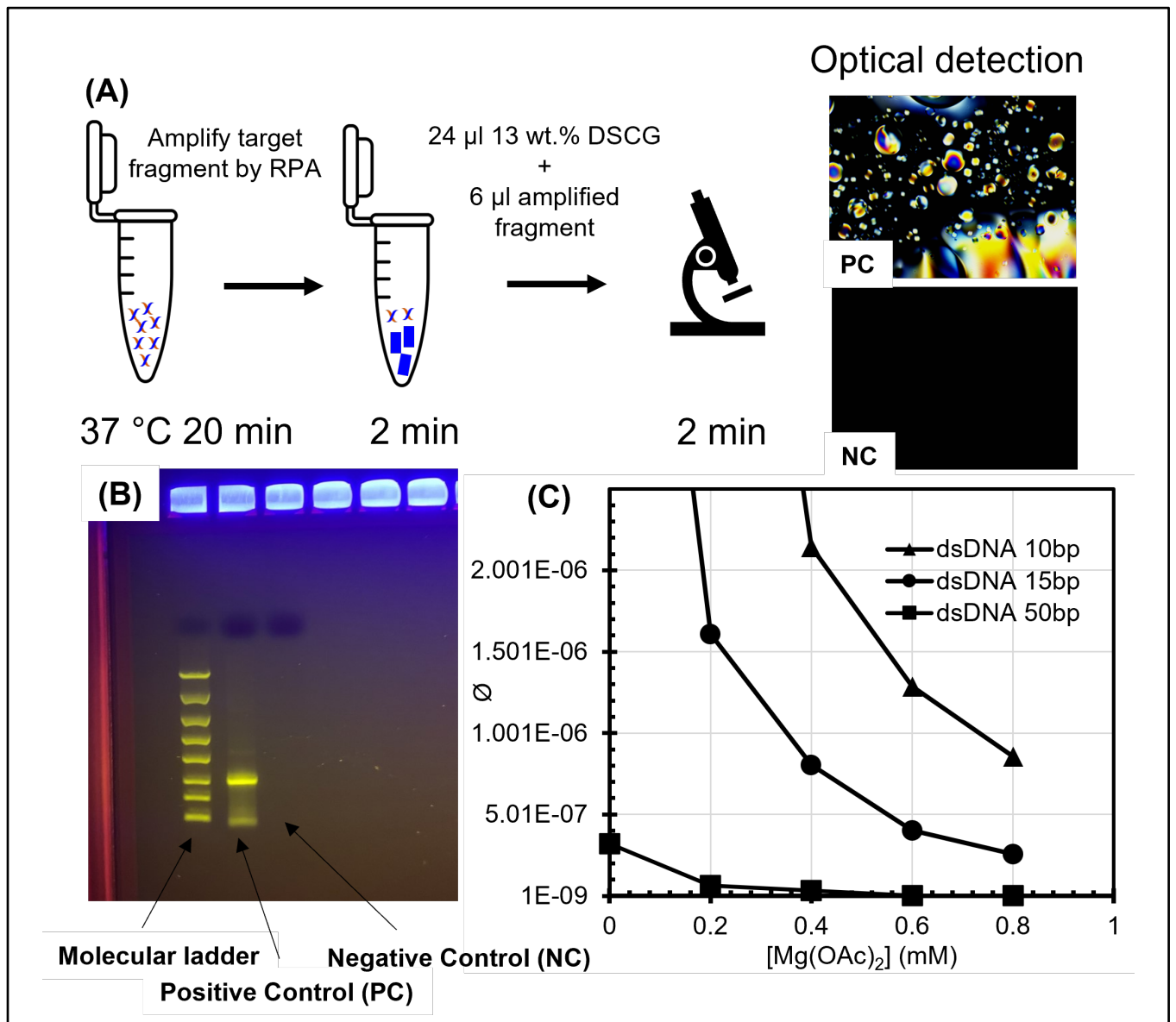

**Figure 7: Rapid detection of amplified dsDNA.** (A) The workflow of the overall process for LCLC rapid detection of amplified RPA samples within 25 min. B) Photograph of gel electrophoresis of cleaned positive and negative samples. C) Role of $Mg(OAc)_2$ on dsDNA volume fractions needed for first-order phase transitions of DSCG.

Typical concentration of DNA (copies converted to moles) following an RPA reaction (bps) is in the ~nM range. We note that our prior observations (in Figures 1, 2, 4) however required DNA concentration in the ~µM range to induce a phase change in DSCG. We note that $Mg(OAc)_2$ present within the RPA mixture can influence the phase behavior of DSCG. Previous work has hypothesized that bound $Mg^{2+}$ can create bridging interactions between neighboring aggregates and therefore a larger propensity to form condensed phases, including precipitation at higher concentrations of DSCG (*60*). To test this hypothesis, we added $Mg(OAc)_2$ to our previous ds-DNA samples (from Integrated DNA Technologies (IDT) and studied the phase behavior. Figure 7C presents the effect of $Mg(OAc)_2$ concentration on the critical DNA concentration required to induce the isotropic-to-biphasic phase transition of DSCG in our system. We found that at a concentration of 0.8 mM or below, $Mg(OAc)_2$ does not precipitate 9.5 wt% DSCG solutions. In the presence of $Mg(OAc)_2$, we observe a significant change in the overall concentrations of dsDNA needed to induce a phase change in DSCG. Remarkably, for 50 bp dsDNA, we observe phase condensation of 9.5 wt% DSCG at concentrations as low as 10 nM, whereas, in the absence of $Mg(OAc)_2$, we observed phase transitions only at 1.5 $\mu$M. The bridging effects induced by $Mg(OAc)_2$ likely result in condensation of DSCG at concentrations of DNA that are in the regime of an RPA-post amplification event thereby opening up avenues for fast optical reporting of the presence of DNA. Interestingly, DNA-mediated chirality transfer persists even at concentrations relevant to post-amplification DNA detection. Figure S8 shows that 5 nM dsDNA in the presence of 0.8 mM $Mg^{2+}$ produces a twist angle of 36.4° in 9.5 wt% DSCG tactoid, demonstrating that nanomolar quantities of DNA are sufficient to transfer chirality and induce condensation of the LC-DSCG phase.

## CONCLUSION

In this work, we demonstrate that chiral depletants fundamentally alter the phase behaviour and structural organization of lyotropic chromonic liquid crystals beyond the predictions of classical depletion frameworks. DNA, both duplex and single-stranded, induce phase condensation of DSCG at volume fractions three orders of magnitude lower than conventional achiral depletants such as PEG, while remaining excluded from the condensed phase. This establishes that phase separation is not governed solely by excluded volume effects but is strongly modulated by the intrinsic chirality of the depletant. Through polarized optical microscopy and twist angle quantification, we show that DNA transmits chirality across phase boundaries, resulting in a macroscopic helical director configuration within the condensed phase. Importantly, this twist is observed in both confined tactoids and extended nematic regions in rectangular capillary, demonstrating that chirality is an intrinsic property of the condensed phase rather than a confinement-induced artefact. Control experiments with achiral PEG confirm that shape factor alone cannot account for the phase condensation behaviour observed at subcritical volume fraction of chiral dsDNA, while results with flexible spherical ssDNA reveal that chirality can dominate over particle shape. Absorption spectroscopy and aggregation modeling reveal that chiral depletants enhance the effective aggregation of DSCG, shifting the equilibrium toward longer stacks that stabilize liquid-crystalline order. This provides a mechanistic link between chirality, aggregate growth, and phase behaviour, suggesting that chiral interactions precondition the system toward ordered states, thereby lowering the energetic barrier for phase transition. Beyond fundamental insights, these findings enable a practical application: an ultrasensitive optical method for detecting DNA amplification. The ability of DSCG to respond to nanomolar DNA concentrations, including tunability, offers a simple, label-free alternative to conventional detection techniques. More broadly, the observed ability of DNA to transmit chirality across phase boundaries while remaining in the continuous phase provides a physical framework for understanding how chiral biomolecules can influence mesoscale organization without direct incorporation into condensed domains. In crowded cellular environments, where macromolecular phase separation is ubiquitous, such long-range chiral fields might influence the orientation and ordering of nearby assemblies, contributing to the emergence of hierarchical organization in biomolecular condensates. Future studies using small-angle X-ray scattering (SAXS) will further provide molecular-level insight into how chiral depletants reorganize the chromonic aggregation landscape Overall, this study establishes chirality as a key governing parameter in depletion-driven phase behavior and liquid crystal self-assembly, opening new directions for understanding biomolecular condensates and designing responsive soft matter systems for sensing and biotechnology.

**ACKNOWLEDGEMENT**

KN acknowledges funding from NSF grant #2443873.

National Institute of Standards and Technology authors were solely funded by the United States Government. Certain commercial firms and trade names are identified in this paper to specify the usage procedures adequately for reproducibility. Such identification is not intended to imply recommendation or endorsement by the National Institute of Standards and Technology, nor is it intended to imply that related products are necessarily the best available for the purpose.

Supporting Information for:

# Depletant-DNA Induces Chirality in the Condensed Domains of Lyotropic Liquid Crystals

**Zeba Afia Hasan, Elizabeth Adeogun, Harold Hatch, Jacob Monroe, Karthik Nayani**

## Materials and Methods

*Materials*

Duplex DNA (dsDNA) oligomers (HPLC purified) of 10, 20, 30, and 50 base pairs in length were purchased from Integrated DNA technologies (San Diego, California, USA) and obtained as lyophilized powder. Oligomers were typically purchased in 250 nmol concentration. To obtain stock solutions of 100 μM and 250 μM dsDNA, the calculated volume of nuclease-free water was added to the lyophilized powder. The experiments were performed with the following self-complementary (SC) oligonucleotides, here reported from 5′ to 3′ end: CGCAATTGCG (10 bp, SC), AACGCAAAGATCTTTGCGTT (20 bp, SC), TACGGATCGCAGCTGGGTTAGGGAAGTTGG (30 bp, SC), CAAGCTGTGCCTTGGGTG GCTTTGGGGCA TGGACA TTGATCCTT ATAAAG-3' (50 bp, SC). Further experiments were performed with fluorescein labeled dsDNA (10 bp). The phase behavior of DSCG in the presence of single-stranded DNA (ssDNA) was investigated using synthetic ssDNA oligonucleotides purchased from Thermo Fisher Scientific (Waltham, MA, USA). Three ssDNA sequences of different lengths were used in this study: a 15-mer (5′-CGGGATCCATGCGAC-3′), a 37-mer (5′-CGGGATCCATGCGACA TCCTTTAGTGATGGGTAACTG-3′), and a 100-mer (5′-CGGGATCCATGCGACATCCTTTAGTGATGGGTAACTGCGGGATCCATG CGACATCCTTTAGTGATGGGTAACTGCGGGATCCATGCGACC-3′). Fluorescein labeled ssDNA (100 bases) with 5′ fluorescein (FAM) modifica was also purchased from Thermo Fisher Scientific (Waltham, Massachusetts, USA). Disodium cromoglycate (DSCG) was purchased from Tokyo Chemical Industry (TCI) America (Portland, Oregon, USA). Stock solutions of 13 wt% DSCG were prepared by dissolving the calculated mass in deionised-distilled water. The mixture was placed on a shaker for 4 hours and heated to 30 °C to ensure homogeneity. In this experiment, we also used polyethylene glycol (PEG) with molecular weight of 2000, 4000, 10000 and 35000 Daltons purchased from Sigma-Aldrich, Inc. (St Louis, Missouri, USA). Fisher's finest premium grade glass slides and cover glass were purchased from Fisher Scientific (Pittsburgh, Pennsylvania, USA). Deionized water with a resistivity of 18.2 MΩ.cm was obtained using a Milli-Q system (by Millipore) and was used in the experiments.

*Liquid Crystal Formulations*

Aqueous mixtures of disodium cromoglycate (DSCG) and DNA were prepared to investigate depletion-induced liquid-liquid crystalline phase separation (LLCPS). Stock solution of DSCG was first prepared in deionized water at concentrations below the nematic threshold of ~13 wt% and subsequently mixed with defined concentrations of DNA to achieve final compositions of interest. The mixtures were gently hand-mixed and equilibrated to ensure homogeneity prior to characterization. The DSCG-DNA system exploits entropic depletion interactions, wherein the exclusion of rod-like DNA from regions of high DSCG concentration results in an osmotic imbalance that promotes condensation of DSCG into a dense, LC-rich phase coexisting with a dilute supernatant. The compositions were systematically varied to probe the onset of isotropic-biphasic-nematic transitions, with particular emphasis on low DNA volume fractions that are sufficient to induce phase separation. Turbidity was observed immediately after adding DNA. All solutions were prepared at room temperature and used directly as a suspension. The LC-rich bottom phase was immediately placed on a glass slide and observed under both bright field and cross-polarized optical microscopy. Optical microscopy images were obtained using an Olympus BX41 fitted with 40X objectives. Polarized and brightfield images were captured in the presence and absence of a polarizer respectively. The size of the coacervate droplets were measured by Image J software.

*Characterization of Duplex DNA-DSCG Mixtures*

The phase identification was performed using cross-polarized optical microscopy. For optical observations, the mixture of DSCG-dsDNA was introduced via capillary action into glass sandwich cells separated by myler spacers and sealed with epoxy glue. Cross-polarized optical microscopy images were obtained using an Olympus BX41 microscope fitted with 10X objectives. Temperature dependence studies were performed on a hot stage, heating from biphasic/nematic to isotropic was done at 0.2 °C/min. 89 4.4.3.

*Fluorescence Microscopy*

To observe the phase distribution of DSCG-dsDNA and DSCG-ssDNA mixtures, we used fluorescently labelled DNA. DSCG solution with fluorescein labelled dsDNA and fluorescein labelled ssDNA was observed under an inverted microscope (Olympus CX60).

*Recombinase Polymerase Amplification*

RPA reactions were carried out using the commercially available RPA reagent kits, provided in the TwistAmp Liquid Basic kit, available from TwistDX Ltd (Cambridge, UK). The reactions were carried out according to the manufacturer's recommended protocol, and contained positive control primer, pre-master mix (2x reaction buffer, dNTPs, water, 10x Basic E-mix), 20x core reaction mix, magnesium acetate and the template; final volume was 50 μL. The final volume for the positive control (PC) sample included 1 μL of 289 bp dsDNA template with a concentration of 1000 copies/μL, while the negative control (NC) sample included an extra 1 μL of water. The reactions were incubated at 37°C for 20 min. Real-time detection of amplified RPA reactions incorporating LCLC was explored by directly adding calculated volumes of amplified DNA to a specified volume and concentrations of DSCG.

*UV-Vis Absorption Spectroscopy*

UV-vis absorption spectra were measured using a BioTek Epoch spectrophotometer in the 200-500 nm range and a 96-well plate corresponding to a path length of $d = 0.322$ cm. The absorbance of dilute solutions of 1 mM DSCG, 1 mM DSCG doped with 5 μM PEG, dsDNA, and ssDNA were measured and used to determine the absorption coefficients of the complex coacervate phase. UV-vis measurements were conducted in triplicate. A lower DSCG concentration (1 mM) was employed for the absorbance measurements, as opposed to the ~10 wt% concentration used in the coacervation experiments, as the high DSCG concentration caused an oversaturation of the signal.

*Theoretical Considerations for Analyzing Length Distribution of DSCG Aggregates*

To interpret the UV-vis spectra, we employed exciton coupling theory, which describes the electronic interactions between neighbouring π -π stacked DSCG molecules during aggregation. The model assumes that the dominant spectral changes arise from nearest neighbour electronic coupling, while the fundamental stacking geometry of DSCG remains unchanged during aggregate growth. Previous studies showed that the largest spectral change occurs during dimer formation, followed by progressively smaller changes as aggregates elongate. Under these assumptions, the absorption coefficient of an aggregate containing N molecules is approximated as

$a_N = a_1 + 2\beta \cos(\frac{\pi}{N+1})$ where $a_N$ is the absorption coefficient of an aggregate of size N, $a_1$ is the absorption coefficient of isolated DSCG molecules, and β is the excitonic coupling constant. As N increases, the absorption coefficient decreases toward $a_1 + 2\beta$. Thus, reductions in the molar absorption coefficient indicate an increase in average aggregate length without disrupting the native π–π stacked chromonic architecture.

To estimate aggregate size distributions from the absorption measurements, we modelled DSCG self-assembly using an isodesmic law-of-mass-action framework. The model assumes: (i) reversible stepwise aggregation, (ii) identical aggregation energy for each monomer addition, (iii) linear unbranched aggregates, and (iv) thermodynamic equilibrium. The relative volume fraction $X_N$ of aggregates containing N molecules is given by $X_N = N(X_1 \exp(\frac{E}{k_BT})^N \exp(-\frac{E}{k_BT}))$, with $X_1 = \frac{(1+2\phi\exp(E/k_BT)) - \sqrt{1+4\phi\exp(E/k_BT)}}{2\phi\exp(2E/k_BT)}$, $\phi = \frac{cM}{cM+\rho}$. Within this framework, the aggregate size distribution depends sensitively on both concentration and aggregation energy. Because the aggregation probability scales exponentially with $\bar{E}$, even modest increases in aggregation energy can strongly bias the equilibrium toward larger aggregation numbers $N$. Lastly, to characterize the length of the aggregate ensemble, we computed the weighted aspect-ratio average $\langle X_N L/D \rangle$. According to the Onsager theory, nematic ordering emerges when the orientational entropy gained from alignment compensates for the translational entropy loss associated with ordering. For rod-like systems, the excluded volume between aligned rods is substantially lower than that between randomly oriented rods, making longer aggregates increasingly favorable for nematic phase formation. Thus, increases in $\langle X_N L/D \rangle$directly indicate enhanced susceptibility toward liquid-crystalline ordering.

*Twist Angle Measurement*

The twist angles were measured following the wave-guiding experiment. The physical idea behind a waveguide is that when white light enters a twisted nematic liquid crystal under Mauguin conditions, the LC acts like a waveguide for polarization. As a result, the direction of polarization of light is continuously guided to follow the local orientation of the liquid crystal director as it twists through the sample. To satisfy Mauguin condition, that is the optical retardation, $\Gamma = \frac{2\pi\Delta n d}{\lambda} \gg$ twist angle, $\tau$ the liquid crystal cell between a polarizer and an analyzer. The polarizer is then rotated every 10°, and at every polarizer orientation, the analyzer was also rotated every 10° from which the minimum intensity position is found. This angle corresponds to the condition where the incident polarization is aligned with the director at one boundary, and the analyzer is perpendicular to the output polarization at the opposite boundary. Since we used rectangular capillaries for our wave-guiding experiment, the minimum transmitted intensities occur when the polarizer is horizontal. The polarizer is then kept fixed at the horizontal position, and the analyzer is rotated to find the minimum intensity. This new analyzer angle gives the twist angle as a deviation from 90°. Twist angle, as per our experiments, was dependent on the depletants' concentrations. As we moved up from 1.95 μM DNA to 3 μM and 5 μM, the twist angle increased from 10° to 20° and finally 60°.

*Calculations of Depletion Interactions Between Plates*

We follow the same procedure for depletion between parallel flat plates as the seminal work of Asakura and Oosawa (1) but explicitly consider hard rods of any aspect ratio. Hard rods are capped cylinders and are thus defined by a length $L$ and diameter $D$ such that half of their total length is $p = \frac{L+D}{2}$. Plates are assumed infinitely thin and edge effects are neglected, which stems from an assumption that the plate area $A$ and total volume accessible to the rod $V$ are both much larger than any rod dimension and the plate separation $h$. Only single depletants are considered which is appropriate in the infinite dilution limit. Increasing the concentration in this limit linearly scales the depletion potential, which is not relevant to the calculations presented.

The depletion potential is given by the Helmholtz free energy $F$ at a fixed plate separation by $\beta F(h) = -\ln Z(h)$, where $Z(h)$ is the canonical partition function at a specific plate separation. To determine $Z(h)$, we integrate over all possible configurations of the rod that do not overlap with the plates. To determine this integral, we divide the total volume into two regions – one within $p$ of any plate, where overlaps are possible depending on the angle of the rod, and one outside this range. We consider a single face of a plate, defining a coordinate system for the rod in terms of its position and orientation relative to the plate, with $z$ being the distance from the rod's center to the plate and $\theta$ the polar angle between the rod's axis and the normal vector of the plate. The other degrees of freedom, namely the translational coordinates in the plane of the plate and the azimuthal angle, may be integrated out to yield a factor of $2\pi A$. The contribution for the region where $z < p$, integrated up to some distance $z'$ away from the surface, is

$$I(z') = 2\pi A \int_0^{\pi} \int_{q(\theta)}^{z'} dz \sin\theta \, d\theta$$

The function $q(\theta) = \frac{1}{2}(L\cos\theta + D)$ is the nearest approach distance possible for a rod at an angle $\theta$ to the plate. We can immediately integrate over $z$ and recognize that the integral over $\theta$ is symmetric over the intervals $\left[0, \frac{\pi}{2}\right]$ and $\left[\frac{\pi}{2}, \pi\right]$, resulting in

$$I(z') = 4\pi A \int_{\theta^*}^{\frac{\pi}{2}} \left(z' - q(\theta)\right) \sin\theta \, d\theta$$

Note that, if $z' < q(\theta)$, an overlap occurs and the contribution to the integral should be zero. To prevent this, we set our lower integration bound to $\theta^* = \cos^{-1}\left(\frac{2z'-D}{L}\right)$, which comes from solving $z' = q(\theta^*)$. This is the smallest angle for which no overlaps occur at a given $z'$. Completing the integral with this lower bound yields

$$I(z') = \frac{4\pi A \left(z' - \frac{D}{2}\right)^2}{L}$$

We are now in a position to write the partition function $Z(h)$ as a piecewise function

$$Z(h) = \begin{cases} 4\pi V - 16\pi A p + 4I(p), & h \geq 2p \\ 4\pi V - 8\pi A p - 4\pi A h + 2I(p) + 2I\left(\frac{h}{2}\right), & D \leq h < 2p \\ 4\pi V - 8\pi A p - 4\pi A h + 2I(p), & h < D \end{cases}$$

The first condition holds when the depletant fits unrestrained between the plates. In the top line, the volume considered by the integral $I(z')$ is subtracted from the total volume before adding the integral. The factor of 4 accounts for the 4 accessible faces of the plates. In the second condition, the rod fits between the two plates and can rotate to some degree, while in the last condition, the rod cannot fit between the two plates even when parallel to their faces.

*Monte Carlo Simulations to Determine Potentials of Mean Force*

Simulations are conducted with the software FEASST. DSCG is represented by hard spherocylindrical rods of length 4.08 nm and diameter 1.0 nm. Two DSCG rods are placed in a parallel orientation at a fixed separation $h$ for each simulation. For each depletant of length $L$ and diameter $D$, 20 simulations are run at equal intervals of the reciprocal separation from DSCG contact ($h = 2$) to the separation at which the depletion interaction becomes a constant ($h = 2 + L + D$). All potentials of mean force are shifted by this constant so that the furthest value shown in the figures is zero. At each DSCG separation, 5 million Widom insertions of a single depletant at uniformly distributed random positions and orientations are attempted. Cubic simulation boxes of edge length 42.004 nm are used for all depletants so that the simulation box is large enough to avoid periodic interactions between even the largest depletants and multiple DSCG rods. The results of each Widom insertion, indexed as $i$, are stored as $w_i = e^{-\beta U_i}$, where $\beta = \frac{1}{k_B T}$ is the reciprocal temperature, $k_B$ is Boltzmann's constant, and the potential energy $U$ is infinity if any particles overlap and zero otherwise. An average is taken over all insertion attempts at a fixed separation so that $w(h) = \langle e^{-\beta U} \rangle_h$. The subscript $h$ on the ensemble average indicates that this is over all insertions at the same separation distance. Potentials of mean force are then computed as $\beta F(h) = -\ln w(h)$.

*Calculations of radius of gyration, aspect ratio and volume fraction*

<u>Polyethylene glycol (PEG)</u>

The radius of gyration ($R_g$) of polyethylene glycol (PEG) was calculated from its molecular weight using the empirical relation (4)

$$R_g = 0.020 M_w^{0.583}$$

where $R_g$is expressed in nanometers and $M_w$is the molecular weight (Da).

For 10 kDa PEG, Mw = 10,000 Da, $R_g$ = 0.02 × (10,000)0.583

= 0.02(214.78)

=4.29 nm

<u>Duplex DNA</u>

Because the duplex DNA oligomers used in this study are significantly shorter than the persistence length of dsDNA (~ 50 nm), they were approximated as rigid cylinders.

The contour length was calculated from $L = N_{bp} \times 0.34$

where $N_{bp}$is the number of base pairs, and 0.34 nm is the rise per base pair.

For 30 bp dsDNA, L = 30 × 0.34 nm = 10.2 nm

Using a duplex diameter, $D = 2.0$ nm,

the aspect ratio becomes, $Aspect\ ratio = \frac{L}{D} = \frac{10.2}{2}\ nm = 5.1\ nm$

Single-stranded DNA

Single-stranded DNA was treated as a flexible polymer chain.
The contour length was calculated from, $L = N_b l_n$
where $N_b$ is the number of nucleotides, and $l_n$ = 0.59nm per nucleotide.
For 100 bases, L = 100 × 0.59 nm = 59 nm

The radius of gyration of an ideal flexible polymer is (5) $<R_g^2> = \frac{N(2L_p)^2}{12} = \frac{LL_p}{6}$
where $N = \frac{L}{2L_p}$ and $L_p$ = 2 nm.
Therefore, $N = \frac{59}{2(2)} = 14.75$
The radius of gyration becomes $\sqrt{\frac{L.L_P}{6}} = \sqrt{\frac{59\times2}{6}} = \sqrt{19.67} = 4.43\ nm$
All the other radius of gyration for PEG and ssDNA and aspect ratio for dsDNA were calculated following the same method.
The volume fraction of each depletant was calculated from its molar concentration and effective molecular volume according to $\phi = nv_p$. Here $n$ is the molecular number density and $v_p$ is the effective volume occupied by one depletant molecule. The molecular number density is related to the molar concentration, *C*, by $n = CN_A \times 10^3$ where C is expressed in mol $L^{-1}$, $N_A = 6.022 \times 10^{23}$ $mol^{-1}$ is Avogadro's constant, and the factor $10^3$ converts liters to cubic meters. Therefore, the volume fraction was calculated directly using $\phi = CN_A \times 10^3 \times v_p$. PEG and ssDNA were approximated as spherical polymer coils, whereas dsDNA was modeled as a spherocylinder.

Volume fraction of 10 kDa PEG

PEG was approximated as a spherical polymer coil with an effective volume given by $v_{PEG} = \frac{4}{3}\pi R_g^3$

For 10 kDa PEG, $R_g = 4.29\ nm$ or $R_g = 4.29 \times 10^{-9} m$

Therefore, the effective volume of one PEG coil was $v_{PEG} = \frac{4}{3}\pi(4.29 \times 10^{-9})^3$

$$v_{PEG} = 3.307 \times 10^{-25} m^3$$

Using the PEG concentration from our experiments, $C_{PEG} = 82\ \mu M = 82 \times 10^{-6} molL^{-1}$

the volume fraction was calculated as $\phi_{PEG} = C_{PEG} N_A \times 10^3 \times v_{PEG}$

$$\phi_{PEG} = (82 \times 10^{-6})(6.022 \times 10^{23})(10^3)(3.307 \times 10^{-25})$$

$$\phi_{PEG} = 1.63 \times 10^{-2}$$

Thus, $\phi_{PEG} = 1.63 \times 10^{-2}$

Volume fraction of 30 bp dsDNA

The 30 bp dsDNA molecule was approximated as a spherocylinder consisting of a cylindrical body and two hemispherical end caps. Its effective volume was calculated from $v_{dsDNA} = \frac{\pi}{4} LD^2 + \frac{\pi}{6} D^3$ where *L* is the contour length and *D* is the dsDNA diameter.

For 30 bp dsDNA, $L = 30 \times 0.34 = 10.2\ nm = 10.2 \times 10^{-9} m$

and $D = 2.0\ nm = 2.0 \times 10^{-9} m$

The cylindrical contribution to the molecular volume was $v_{cylinder} = \frac{\pi}{4} LD^2$

$$v_{cylinder} = \frac{\pi}{4}(10.2 \times 10^{-9})(2.0 \times 10^{-9})^2$$

$$v_{cylinder} = 3.204 \times 10^{-26} m^3$$

The combined volume of the two hemispherical end caps was $v_{caps} = \frac{\pi}{6} D^3$

$$v_{caps} = \frac{\pi}{6}(2.0 \times 10^{-9})^3$$

$$v_{caps} = 4.189 \times 10^{-27} m^3$$

Therefore, the total effective volume of one dsDNA molecule was

$$v_{dsDNA} = v_{cylinder} + v_{caps}$$

$$v_{dsDNA} = 3.204 \times 10^{-26} + 4.189 \times 10^{-27}$$

$$v_{dsDNA} = 3.623 \times 10^{-26} m^3$$

Using the 30 bp dsDNA concentration from our experiments, $C_{dsDNA} = 0.14\ \mu M = 0.14 \times 10^{-6} mol L^{-1}$

the volume fraction was calculated as $\phi_{dsDNA} = C_{dsDNA} N_A \times 10^3 \times v_{dsDNA}$

$$\phi_{dsDNA} = (0.14 \times 10^{-6})(6.022 \times 10^{23})(10^3)(3.623 \times 10^{-26})$$

$$\phi_{dsDNA} = 3.09 \times 10^{-6}$$

Thus, $\phi_{dsDNA} = 3.09 \times 10^{-6}$

Volume fraction of 100 bases ssDNA

The 100-base ssDNA molecule was approximated as an effective spherical polymer coil. Its effective molecular volume was calculated from $v_{ssDNA} = \frac{4}{3}\pi R_g^3$

Using the calculated radius of gyration, $R_g = 4.44\ nm = 4.44 \times 10^{-9} m$

the effective volume of one ssDNA coil was $v_{ssDNA} = \frac{4}{3}\pi(4.44 \times 10^{-9})^3$

$$v_{ssDNA} = 3.666 \times 10^{-25} m^3$$

Using the 100 bases ssDNA concentration from our experiments, $C_{ssDNA} = 0.3\ \mu M = 0.3 \times 10^{-6} mol L^{-1}$

the volume fraction was calculated as $\phi_{ssDNA} = C_{ssDNA} N_A \times 10^3 \times v_{ssDNA}$

$$\phi_{ssDNA} = (0.3 \times 10^{-6})(6.022 \times 10^{23})(10^3)(3.666 \times 10^{-25})$$

$$\phi_{ssDNA} = 6.29 \times 10^{-5} \approx 6.3 \times 10^{-5}$$

Thus, $\phi_{ssDNA} = 6.3 \times 10^{-5}$

All the other volume fractions for PEG, dsDNA, and ssDNA were calculated following the same method.

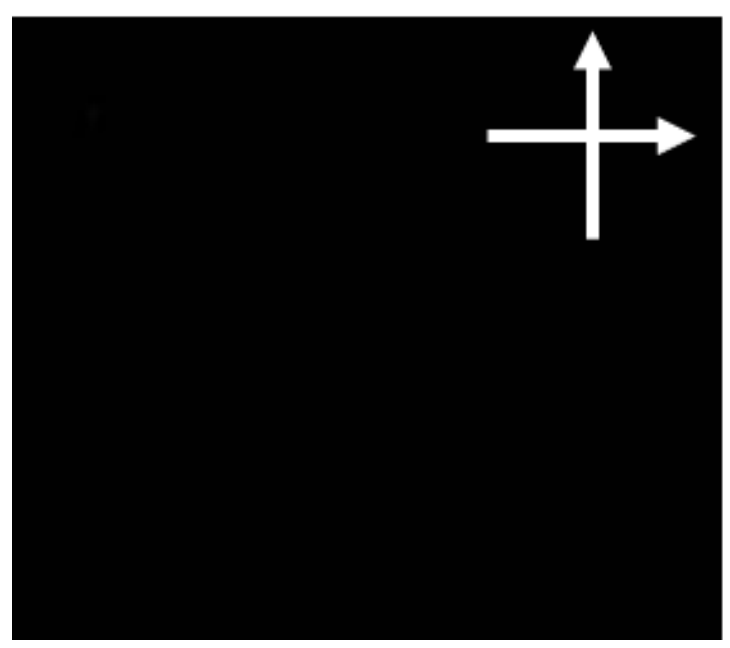

**Figure S1: An image of the isotropic phase of DSCG under crossed-polarizers configuration.**

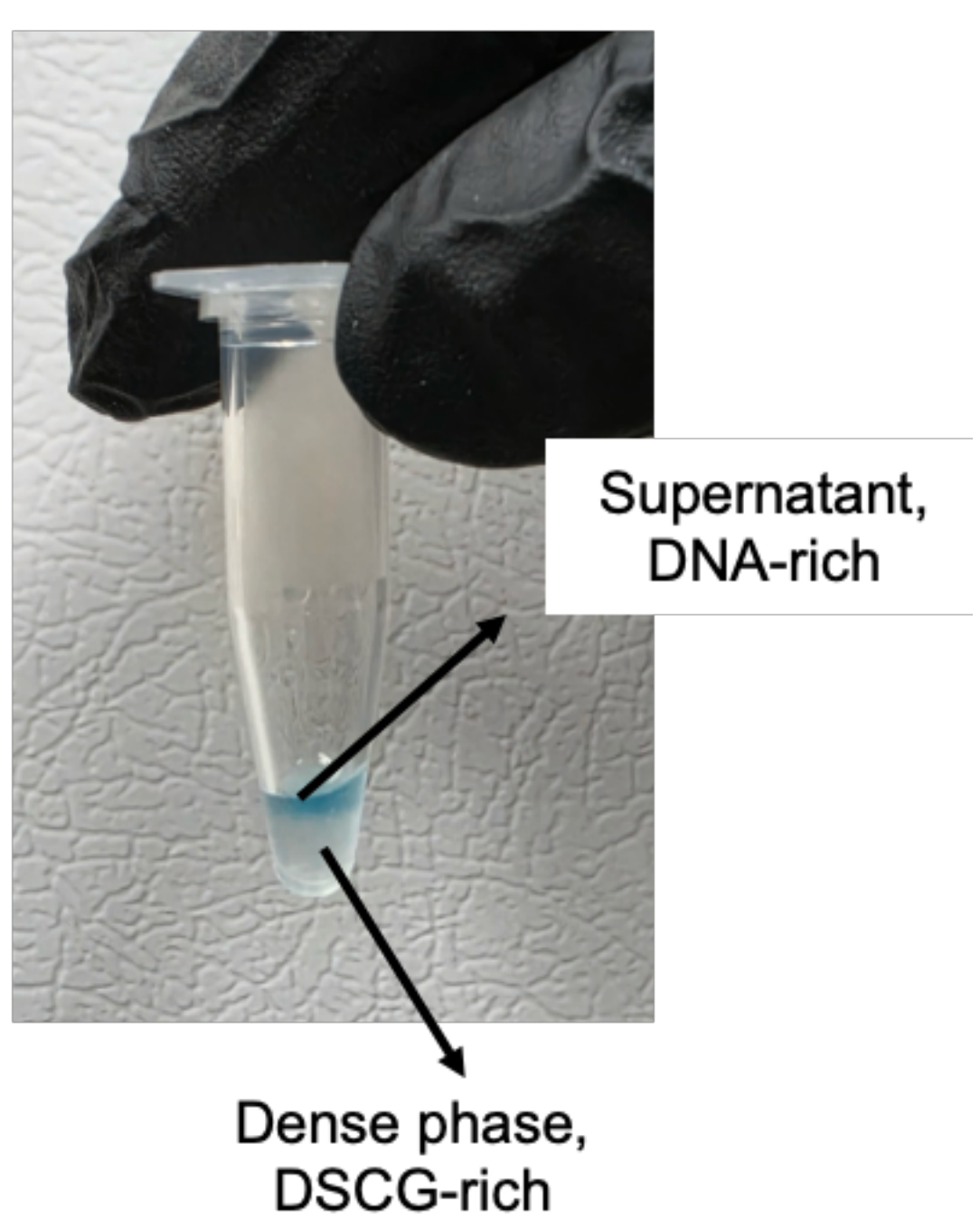

**Figure S2: Macroscopic phase separation of a sample containing 9.5 wt% DSCG in presence of fluorescently tagged 10 bp dsDNA at Ø = $2\times10^{-4}$**

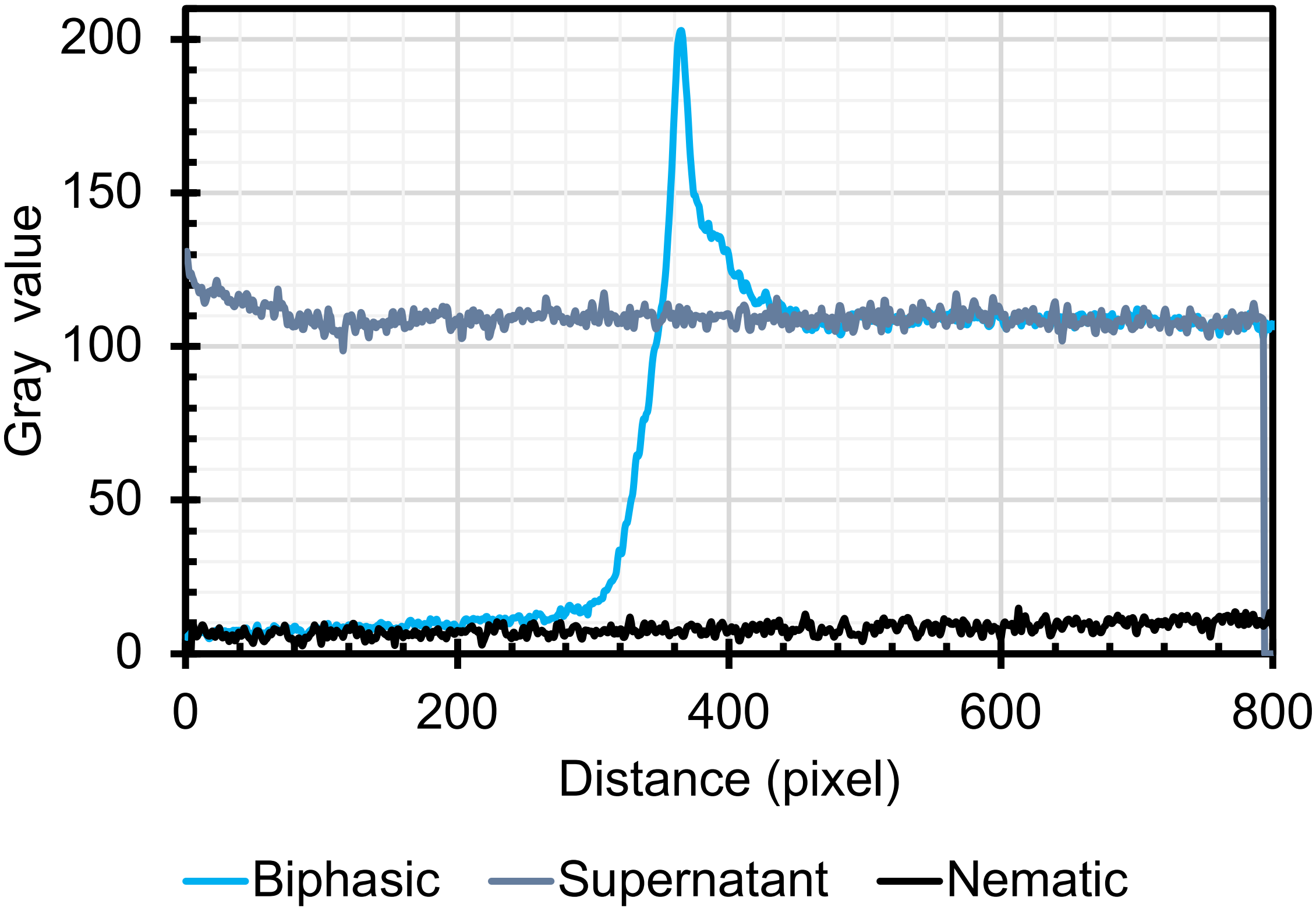

**Figure S3: Line scan intensity profile for biphasic, supernatant, and nematic phases presented in Figure 2.**

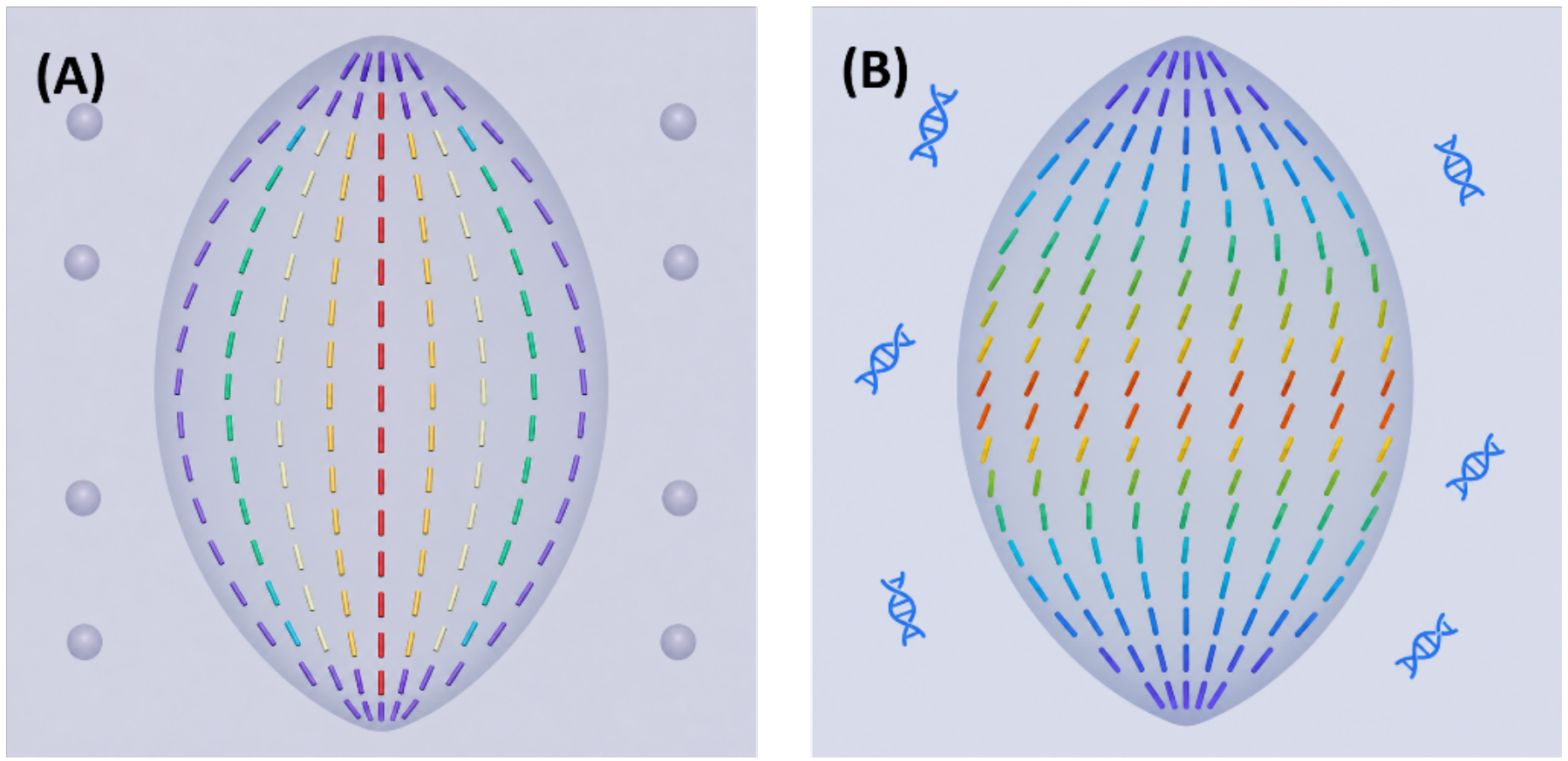

**Figure S4: A schematic of the director configuration inside a DSCG tactoid.** A) Tactoid nucleated upon addition of achiral PEG. B) Tactoid nucleated upon introducing dsDNA, showing a 12.12º twist in the center.

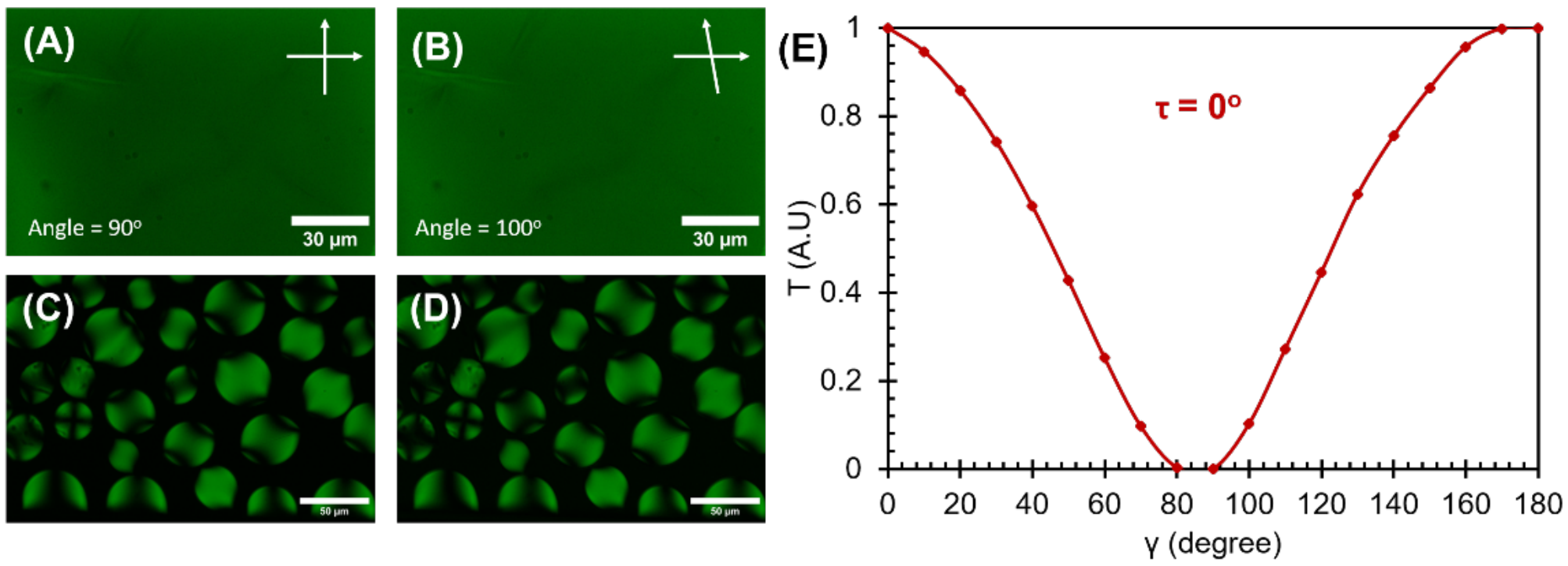

**Figure S5: Twist angle quantification with PEG 35 kDa.** A) crossed-POM micrograph of 9.5 wt% DSCG + 300 µM PEG 35 kDa in rectangular capillary. B) slightly uncrossed-POM of 9.5 wt% DSCG + 300 µM PEG 35 kDa. C) crossed-POM of a DSCG tactoid in 9.5 wt% DSCG + 300 µM PEG 35 kDa system. D) slightly uncrossed-POM micrograph of that tactoid in 9.5 wt% DSCG + 300 µM PEG 35 kDa system. E) Light intensity transmitted through the center of the sample as a function of the angle difference between the polarizer and analyzer. Red trendlines represent data for both rectangular capillary and tactoid in 9.5 wt% DSCG + 300 µM PEG 35 kDa system. Trendlines for both systems overlap due to both showing transmittance minima at 90°.

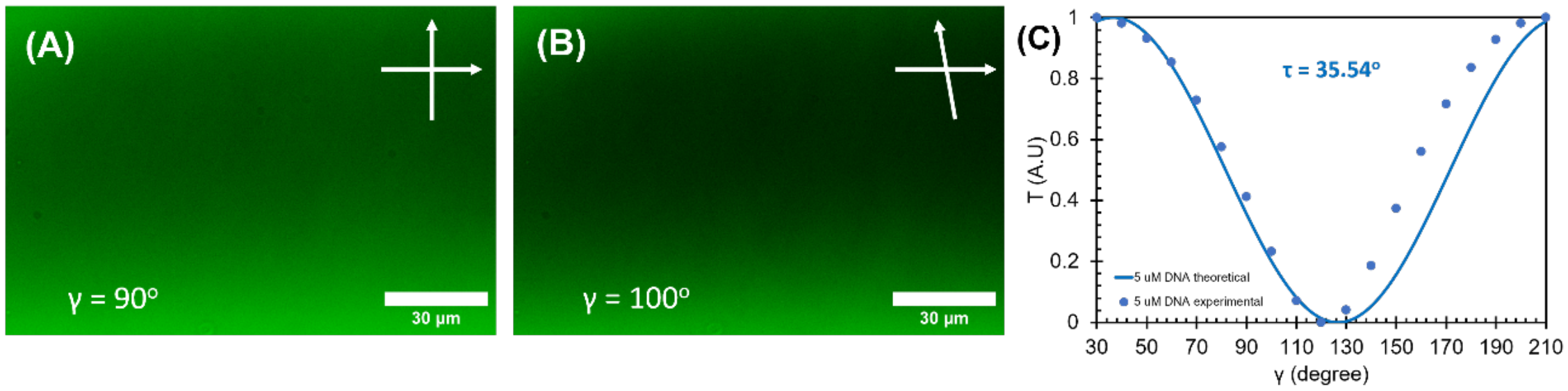

**Figure S6: Twist angle quantification with 5 µM dsDNA.** A) crossed-POM micrograph of 9.5 wt% DSCG + 5 µM dsDNA. B) slightly uncrossed-POM micrograph of 9.5 wt% DSCG + 5 µM dsDNA. C) Light intensity transmitted through the center of the sample as a function of the angle difference between the polarizer and analyzer. Data points represent the experimental data and trendlines represent the fit to equation 1.

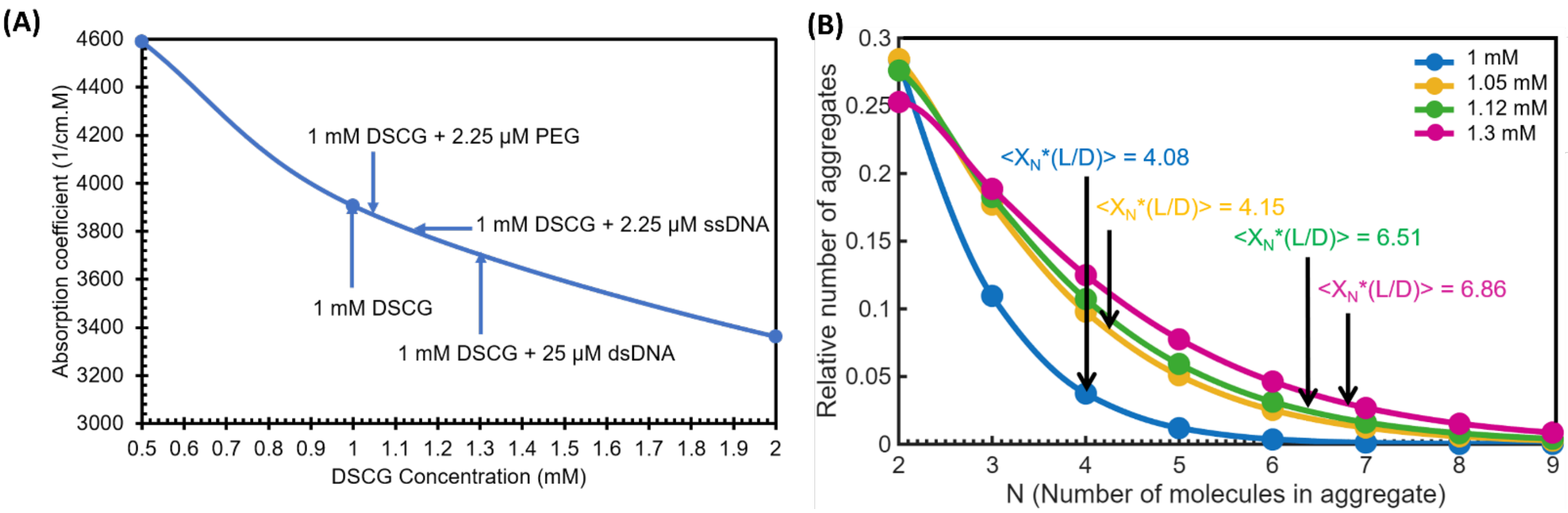

**Figure S7: Chirality-enhanced aggregation energetics promote formation of longer DSCG aggregates.** A) Calibration curve for deriving effective DSCG concentration corresponding to DSCG with PEG, dsDNA and ssDNA depletants. B) Aggregate length distributions calculated using the isodesmic aggregation model with concentration-dependent aggregation energies. Increasing effective aggregation energy shifts the distribution toward larger aggregation numbers $N$, producing higher ensemble-averaged aspect ratios $\langle X_N \frac{L}{D} \rangle$, indicative of longer and more correlated chromonic aggregates.

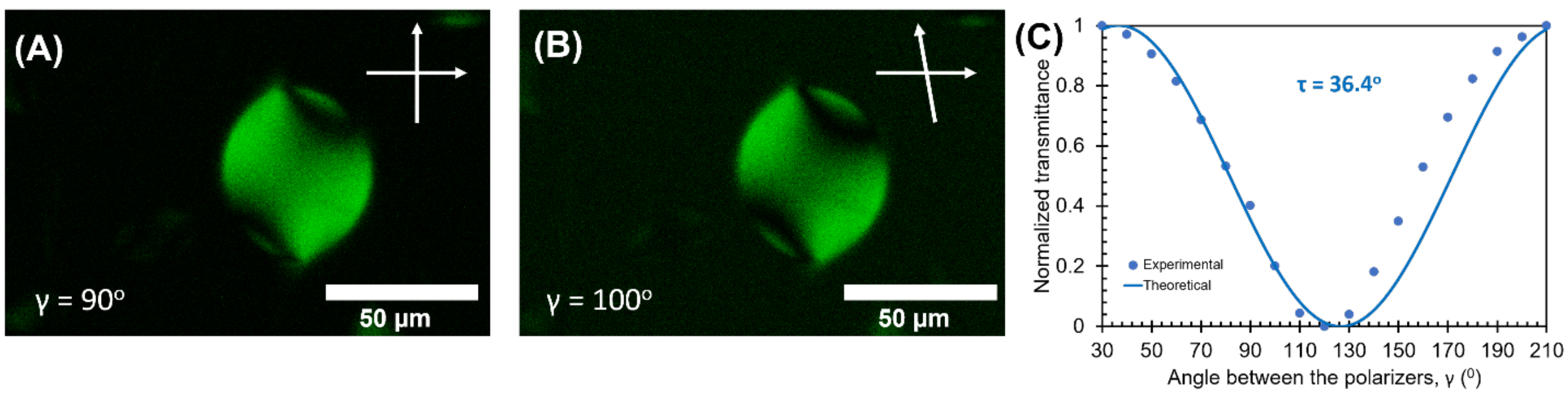

**Figure S8: Twist angle quantification with nanomolar dsDNA.** A) crossed-POM micrograph of 9.5 wt% DSCG tactoid nucleated in the presence of 5 nM dsDNA (30 bp) and 0.8 mM $Mg^{2+}$ salt. B) slightly uncrossed-POM micrograph of that 9.5 wt% DSCG tactoid nucleated in the presence of 5 nM dsDNA (30 bp) and 0.8 mM $Mg^{2+}$ salt. C) Light intensity transmitted through the center of the sample as a function of the angle difference between the polarizer and analyzer. Data points represent the experimental data and trendlines represent the fit to equation 1.

**Disclaimer**: Certain commercial firms and trade names are identified in this paper to specify the usage procedures adequately for reproducibility. Such identification is not intended to imply recommendation or endorsement by the National Institute of Standards and Technology, nor is it intended to imply that related products are necessarily the best available for the purpose.